\documentclass[journal,twocolumn,10pt]{IEEEtran}
\usepackage[T1]{fontenc}
\usepackage{cite}
\usepackage[cmex10]{amsmath}
\usepackage{amsthm}
\usepackage{amssymb}
\usepackage{mathtools}

\usepackage{algorithmic}
\usepackage{textcomp}
\usepackage[usenames, dvipsnames]{color}
\usepackage[linesnumbered,ruled]{algorithm2e}
\usepackage{subcaption}
\usepackage{graphicx}  
\usepackage{url}

\newtheorem{theorem}{Theorem}
\newtheorem{proposition}{Proposition}

\DeclareMathOperator*{\argmax}{argmax} 

\ifCLASSINFOpdf
\else
\fi

\begin{document}
%
\title{Supervised and Semi-Supervised Learning for MIMO Blind Detection with Low-Resolution ADCs}
%
%
%

\author{Ly~V.~Nguyen,
	Duy~T.~Ngo,
	Nghi~H.~Tran, A.~Lee~Swindlehurst, and Duy~H.~N.~Nguyen
	\thanks{Ly V. Nguyen and Duy H. N. Nguyen are with the Department of Electrical and Computer Engineering and the Computational Science Research Center, San Diego State University, San Diego, CA, USA  92182 (e-mail: \{vnguyen6, duy.nguyen\}@sdsu.edu).}
	\thanks{Duy T. Ngo is with the School of Electrical Engineering and Computing, The University of Newcastle, Callaghan, NSW 2308, Australia (e-mail: duy.ngo@newcastle.edu.au).}
	\thanks{Nghi H. Tran is with the Department of Electrical and Computer Engineering, The University of Akron, Akron, Ohio, USA (e-mail: nghi.tran@uakron.edu).}
	\thanks{A. Lee Swindlehurst is Department of Electrical Engineering and
		Computer Science, Henry Samueli School of Engineering, University of
		California, Irvine, CA, USA 92697 (e-mail: swindle@uci.edu).}}

%
%

\markboth{Submitted for Journal Publication}%
{Ly V. Nguyen \MakeLowercase{\textit{et al.}}}
%



\maketitle

\begin{abstract}
The use of low-resolution analog-to-digital converters (ADCs) is considered to be an effective technique to reduce the power consumption and hardware complexity of wireless transceivers. However, in systems with low-resolution ADCs, obtaining channel state information (CSI) is difficult due to significant distortions in the received signals. The primary motivation of this paper is to show that learning techniques can mitigate the impact of CSI unavailability. We study the blind detection problem in multiple-input-multiple-output (MIMO) systems with low-resolution ADCs using learning approaches. Two methods, which employ a sequence of pilot symbol vectors as the initial training data, are proposed. The first method exploits the use of a cyclic redundancy check (CRC) to obtain more training data, which helps improve the detection accuracy. The second method is based on the perspective that the to-be-decoded data can itself assist the learning process, so no further training information is required except the pilot sequence. For the case of 1-bit ADCs, we provide a performance analysis of the vector error rate for the proposed methods. Based on the analytical results, a criterion for designing transmitted signals is also presented. Simulation results show that the proposed methods outperform existing techniques and are also more robust.
\end{abstract}


%
\IEEEpeerreviewmaketitle

\section{Introduction}
Wireless spectrum is limited and the currently used spectrum, $700$ MHz $-$ $2.6$ GHz, is not sufficient to support the demand of future wireless users \cite{rappaport2013millimeter}. Recently, massive MIMO and millimeter-wave (mmWave) communications have attracted great attention and have been considered promising solutions for this challenge \cite{Boccardi2014Five,Jungnickel2014role,andrews2014will,wang2014cellular}. While massive MIMO communications enhance the throughput by using tens to hundreds of antenna elements \cite{Marzetta2010Noncooperative,Rusek2013scaling,Hoydis2013massive}, mmWave technologies utilize higher frequencies, $30$ GHz $-$ $300$ GHz, where available bandwidths are capable of providing very high communication speed (e.g., on the order of Gbps) \cite{pi2011introduction,rajagopal2011antenna}. 

Although massive MIMO and mmWave technologies are being deployed for next generation wireless networks, they still appear to face many technical challenges. More specifically, in massive MIMO systems, a large number of RF chains are required, resulting in significant increases in hardware complexity, system cost and power consumption \cite{Ji2017Overview}. For mmWave systems, the sampling rate has to be sufficiently high to satisfy the Nyquist theorem, which will lead to high power consumption by the analog-to-digital converters (ADCs) \cite{walden1999analog,Murmann}. In addition, a massive number of active antennas and a high sampling rate demand prohibitively high bandwidth on the fronthaul link between the baseband processing unit and the RF chains. For example, a receiver that is equipped with $100$ antennas, where each antenna employs two separate ADCs for the in-phase and quadrature components, and where each ADC samples at a rate of $5$ GS/s with $10$-bit precision would produce $10$ Terabit/s of data, which is much higher than the rates of the common public radio interface in today's fiber-optical fronthaul links \cite{CPRI2013}.

A promising solution for the these issues is to use low-resolution ADCs (i.e., $1$-$3$ bits precision) since the power consumption of the ADCs increases exponentially with the number of bits per sample and linearly with the sampling rate \cite{walden1999analog,Murmann}. In the extreme case of $1$-bit ADCs, automatic gain controls are not required since the quantization requires only a single comparator for each of the in-phase and quadrature channels, and many other RF components such as mixers, frequency synthesizers and local oscillators can also be eliminated in some system architectures \cite{Donnell2005ultraWideband,Mezghani0217mmWave}. However, channel estimation and data detection are significantly more challenging when low-resolution ADCs are employed due to their strong nonlinear behavior. Numerous detection methods have been proposed in the literature \cite{Choi2015Quantized,Wang2014Convex,Mezghani2008Maximum,choi2016near,Mezghzni2007Modified,Wang2015Multiuser,Xiong2017Low,Jeon2018One,wen2016bayes} to deal with such nonlinearities. Maximum-likelihood (ML) detection approaches are studied in \cite{Choi2015Quantized,Wang2014Convex,Mezghani2008Maximum}. The ML detection problem was relaxed to a convex optimization program in \cite{choi2016near,Wang2014Convex} for it to be solvable by low-complexity algorithms. A zero-forcing (ZF) detector was introduced in \cite{Choi2015Quantized} and minimum mean squared error (MMSE) detectors were proposed in \cite{Mezghzni2007Modified,Wang2015Multiuser}. Several other techniques such as Generalized Approximate Message Passing (GAMP) and sphere decoding were employed in \cite{Xiong2017Low} and \cite{Jeon2018One}, respectively. Bayes inference and the GAMP algorithm were studied in \cite{wen2016bayes} to develop a joint channel-and-data estimation method. 

All of the detection techniques mentioned above are coherent, which means they require Channnel State Information (CSI). However, obtaining CSI in MIMO systems with low-resolution ADCs is difficult due to the significant distortions of the received signals. Different approaches have been proposed to estimate CSI in the presence of low-resolution ADCs. ML channel estimators for $1$-bit ADCs are presented in \cite{choi2016near} and \cite{Mezghani2018Blind}, where the work in \cite{Mezghani2018Blind} focused on sparse broadband channels. The least-squares approach was studied for different scenarios in \cite{Wang2014Convex,Risi2014Massive,Wang2015Multiuser}. The Bussgang decomposition was applied in \cite{li2017channel,Jacobsson2017Throughput} to form the MMSE channel estimator. The mmWave MIMO channel estimation problem was formulated as a compressed-sensing problem in \cite{Mo2018Channel,Rodriguez2016Channel,Rusu2015Adaptive,mo2014channel} by exploiting the sparsity of such channels. Although much progress has been made, the channel estimation accuracy is still severely limited due to the coarse quantization effect of the low-resolution ADCs \cite{Jeon2018supervised}. Longer pilot sequences have been proposed to compensate for the quantization errors, but this often requires sequences that are many times longer than the number of co-channel users \cite{mo2014channel,Mo2018Channel,ivrlac2007mimo}.

Recently, there are several results on blind detection for MIMO systems with low-resolution ADCs reported in \cite{Jeon2018supervised,jeon2017blind,Ly2018learning}. The common approach of these papers was to use a training sequence to learn the nonlinear input-output relations of the system and then perform data detection based on the learned results. Hence, information about the channel is not required. For systems with perfect ADCs, there are also several recent results on blind detection using learning approaches. For example, the blind detection problem was addressed as a clustering problem, which was solved by a deep neural network, the Expectation-Maximization (EM) method, and the K-means clustering technique in \cite{farsad2017detection}, \cite{huang2018machine}, and \cite{liang2016coding}, respectively. Some other works have employed the autoencoder model for end-to-end learning \cite{Dorner2018deep,Oshea2016learning}.

The authors of \cite{Jeon2018supervised,jeon2017blind} proposed three supervised learning methods, referred to as \textit{empirical-Maximum-Likelihood Detection} (eMLD), \textit{Minimum-Mean-Distance Detection} (MMD), and \textit{Minimum-Center-Distance Detection} (MCD). These blind detection methods are simple and easy to implement, but their efficiency is heavily dependent on the training sequence. When the length of the training sequence is short, the learned results do not correctly describe the input-output relations of the system. Based on this observation, we propose in this paper two efficient learning methods to resolve the problem of short training sequences. Since MCD outperforms eMLD and MMD, and the complexity of MCD is also lower than that of eMLD and MMD, we compare our proposed methods to MCD only. Preliminary results on the proposed learning methods were reported in \cite{Ly2018learning}. In this paper, we provide a complete analysis of the proposed methods and make the following contributions:
\begin{itemize}
	\item We propose two learning methods that are capable of achieving more precise input-output relations compared to \cite{jeon2017blind} given the same training sequence, and hence will improve the detection accuracy. The first method exploits the use of the cyclic redundancy check (CRC), should it be available in practical systems, to acquire more data for the training process. The second method is based on the observation that not only the training sequence but the to-be-decoded data also contain useful information about the input-output relationship, and hence can be exploited to improve the learned results.
	\item We show via simulations that the proposed methods are more robust than MCD in terms of the training sequence length. Particularly, for extremely short training sequences, the performance of MCD is degraded significantly while that of our proposed methods is more stable. For example, in a system with $2$ transmit antennas, $16$ receive antennas, and BPSK modulation, the gain in bit error rate (BER) produced by the second proposed method can be up to $7$-$8$ dB for BERs between $10^{-3}$ and $10^{-5}$. Even for moderately long training sequences, the gain of our proposed methods is still considerable, between $3$-dB and $4$-dB.
	\item We provide performance analyses of the vector error rate (VER) for the case of $1$-bit ADCs at both low and high signal-to-noise ratios (SNRs). Assuming perfectly learned input-output relations, we first approximate the pairwise VER at low SNR by using the Bussgang decomposition and use this approximation to derive an upper bound on the VER. The asymptotic VER performance at infinite SNR for Rayleigh fading channels is then analyzed. Simulation results confirm the accuracy of our analyses at both low and high SNRs.
	\item Finally, based on the performance analysis, we propose a criterion for designing transmitted signals when only a portion of all possible signals are used for transmission.
\end{itemize}

\textit{Notation}: Upper-case and lower-case boldface letters denote matrices and column vectors, respectively. The notation $\mathbf{1}$ is a vector where every element is equal to one. $\mathbb{E}[\cdot]$ represents expectation and $\mathbb{P}[\cdot]$ is the probability of some event. Depending on the context, the operator $|.|$ is used to denote the absolute value of a real number, or the cardinality of a set. The transpose and conjugate transpose are denoted by $[\cdot]^T$ and $[\cdot]^H$, respectively. The operator $\operatorname{mod}(a,b)$ calculates $a$ modulo $b$. The notations $\operatorname{Var}[\cdot]$ and $\operatorname{Cov}[\cdot,\cdot]$ denote the variance and covariance, respectively. The integral $\Phi(a) = \frac{1}{\sqrt{2\pi}}\int_{-\infty}^{a}e^{-t^2/2}dt$ is the cumulative distribution function of the standard normal random variable. The notation $\Re\{.\}$ and $\Im\{.\}$ respectively denotes the real and imaginary parts of the complex argument. If $\Re\{.\}$, $\Im\{.\}$ or $\Phi(.)$ are applied to a matrix or vector, they are applied separately to every element of that matrix or vector. 

\section{System Model}
\label{sec_System_model}
\begin{figure*}
	\centering
	\includegraphics[scale=0.9]{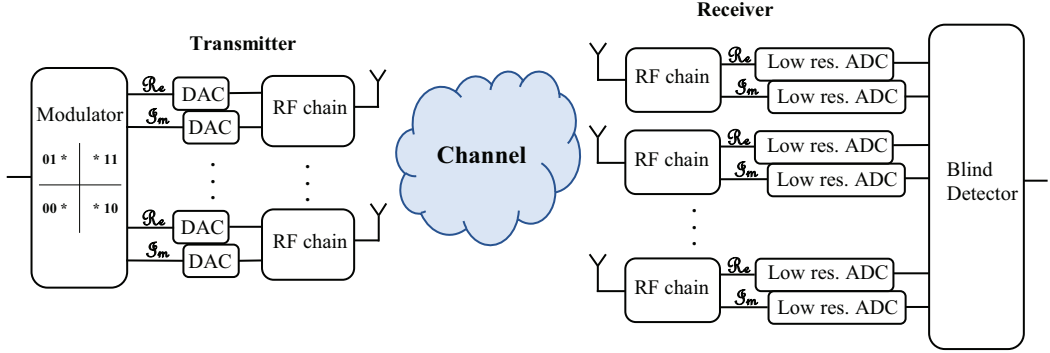}
	\caption{Block diagram of a MIMO communication system with low resolution ADC at the receiver.}
	\label{fig_system_model}
\end{figure*}

The MIMO system we consider, as illustrated in Fig. \ref{fig_system_model}, has $N_\text{t}$ transmit antennas and $N_\text{r}$ receive antennas, where it is assumed that $N_\text{r}\geq N_\text{t}$. Let $\mathbf{x}[n] = [x_1[n], \ldots,x_{N_\text{t}}[n]]^T \in \mathbb{C}^{N_\text{t}}$ be the transmitted signal vector at time slot $n$, where $x_i[n]$ is the symbol transmitted at the $i^{\text{th}}$ transmit antenna. Each symbol $x_i[n]$ is drawn from a constellation $\mathcal{M}$ with a constellation size of $M = |\mathcal{M}|$ under the power constraint $\mathbb{E}[|x_i[n]|^2]=1$. The channel is assumed to be block-fading, and each block-fading interval lasts for $T_\text{b}$ time slots. Hence, the channel $\mathbf{H} = [h_{n_\text{r}n_\text{t}}] \in \mathbb{C}^{N_\text{r} \times N_\text{t}}$ remains constant over $T_\text{b}$ time slots. For the analysis and simulations, we assume a Rayleigh fading channel with independent and identically distributed (i.i.d.) elements and $h_{n_\text{r}n_\text{t}} \sim \mathcal{CN}(0,1)$, but the proposed algorithms are applicable to any channel model. The system model in each block-fading interval is
\begin{equation}
\mathbf{r}[n] = \mathbf{H}\mathbf{x}[n]+\mathbf{z}[n],
\end{equation}
where $\mathbf{r}[n] = [r_1[n],\ldots,r_{N_\text{r}}[n]]^T \in \mathbb{C}^{N_\text{r}}$ is the analog received signal vector, and $\mathbf{z}[n] = [z_1[n],\ldots,z_{N_\text{r}}[n]]^T \in \mathbb{C}^{N_\text{r}}$ is the noise vector. The noise elements are assumed to be i.i.d. with $z_i[n] \sim \mathcal{CN}(0,N_0)$. CSI is unavailable at both the transmitter and receiver sides, i.e., $\mathbf{H}$ is unknown. The signal-to-noise ratio (SNR) is defined as $\rho = N_\text{t}/N_0$.

The considered system employs an ADC that performs $b$-bit uniform scalar quantization, $b\in \{1,2,3,\ldots\}$. The $b$-bit ADC model is characterized by a set of $2^b-1$ thresholds denoted as $\{\tau_1, \ldots, \tau_{2^b-1}\}$. Without loss of generality, we can assume $-\infty = \tau_0< \tau_1 < \ldots < \tau_{2^b-1} < \tau_{2^b} = \infty$. Let $\Delta$ be the step size, so the threshold of a uniform quantizer is given as
\begin{equation}
\tau_l = (-2^{b-1}+l)\Delta, \text{ for } l \in \{1, \ldots, 2^{b}-1\}.
\end{equation}
Let $Q_b(.)$ denote the element-wise quantizer, so that the quantization output is defined as
\begin{equation}
Q_b(r) = \begin{cases}
\tau_l - \frac{\Delta}{2} & \text{if } r \in (\tau_{l-1},\tau_l] \text{ and } l \leq 2^{b}-1,\\
(2^b-1)\frac{\Delta}{2} & \text{if } r \in (\tau_{2^b-1},\tau_{2^b}].
\end{cases}
\end{equation}
It should be noted that this mid-rise uniform quantizer satisfies $Q_b(-r) = -Q_b(r)$ $\forall r$.

The step size $\Delta$ is chosen to minimize the distortion between the quantized and non-quantized signals. The optimal value of $\Delta$ depends on the distribution of the input signals \cite{hui2001asymptotic}. For standard Gaussian signals, the optimal step size $\Delta^\text{standard}_\text{opt}$ can be found numerically as in \cite{dhahir1996uniform}. For non-standard complex Gaussian signals with variance $\sigma^2 \neq 1$, the optimal step size for each real/imaginary signal component can be computed as $\Delta_\text{opt} = \sqrt{\sigma^2/2}\Delta^\text{standard}_\text{opt}$. Hence, the optimal step size in our system is $\Delta_\text{opt} = \sqrt{(N_\text{t}+N_0)/2}\Delta^\text{standard}_\text{opt}$. The variance of the analog received signals $N_\text{t}+N_0$ is assumed to be known at the receiver.

The real and imaginary parts of each received symbol are applied to two separate ADCs. Hence, if $\mathbf{y}[n] = \big[y_1[n],\ldots, y_{N_\text{r}}[n]\big]^T \in \mathbb{C}^{N_\text{r}}$ is the quantized version of the received signal vector $\mathbf{r}[n]$, then $\mathbf{y}[n] = Q_b(\mathbf{r}[n])$ in which $\Re\{y_i[n]\}=Q_b(\Re\{r_i[n]\})$ and $\Im\{y_i[n]\}=Q_b(\Im\{r_i[n]\})$ for all $i \in \mathcal{N}_{\text{r}} = \{1,\ldots,N_\text{r}\}$.

\section{Blind Detection Problem}
This section describes the blind detection problem for the block-fading channel. The first $T_\text{t}$ time slots of each block fading interval contain the training symbol sequence while the remaining $T_\text{d} = T_\text{b} - T_\text{t}$ time slots comprise the data symbol sequence. Let $\check{\mathcal{X}} = \{\check{\mathbf{x}}_1,\check{\mathbf{x}}_2,\ldots,\check{\mathbf{x}}_K\}$ denote the set of all possible transmitted symbol vectors with $K = M^{N_\text{t}}$ and let $\mathcal{K} = \{1,2,\ldots,K\}$. Hereafter, a possible transmitted symbol vector is called a \textit{label}. We first revisit the MCD method presented in \cite{jeon2017blind}, which serves as a baseline for our work. The input-output relations to be learned in the MCD method are $\big \{\mathbb{E}\big[\mathbf{y}|\mathbf{x} = \mathbf{\check{x}}_k\big], k\in \mathcal{K}\big \}$, in which $\mathbb{E}\big[\mathbf{y}|\mathbf{x} = \mathbf{\check{x}}_k\big]$ represents the centroid of the received quantized signal given that the label $\mathbf{\check{x}}_k$ is transmitted. The MCD data detection is given by
\begin{equation}
f(\mathbf{y}[n]) = \underset{k \in \mathcal{K}}{\operatorname{argmin}}  \Big\|\mathbf{y}[n]-\mathbb{E}\big[\mathbf{y}|\mathbf{x} = \mathbf{\check{x}}_k\big]\Big\|_2,
\label{eq_MCD}
\end{equation}
where $\mathbf{y}[n]$ is the received data symbol vector at time slot $n$ with $n \in \{T_\text{t}+1,\ldots,T_\text{b}\}$. Thus, the MCD approach identifies the index of the transmitted label as the one whose centroid is closest to the received vector. Denote $\check{\mathbf{y}}_k = \mathbb{E}\big[\mathbf{y}|\mathbf{x} = \mathbf{\check{x}}_k\big]$; each $\check{\mathbf{y}}_k$ is called a \textit{representative vector} for the label $\check{\mathbf{x}}_k$. There are $K$ representative vectors $\check{\mathcal{Y}} = \{\check{\mathbf{y}}_1,\check{\mathbf{y}}_2,\ldots,\check{\mathbf{y}}_K\}$. Thus, the MCD method has to learn $\check{\mathcal{Y}}$ in order to perform the detection task. We now present two MCD training methods from \cite{jeon2017blind,Ly2018learning,Jeon2018supervised} that help the receiver empirically learn $\check{\mathcal{Y}}$. 
\subsection{Full-space Training Method}
Since the transmitted signal space $\check{\mathcal{X}}$ contains $K$ labels, a straightforward method to help the receiver learn $\check{\mathcal{Y}}$ is using a training sequence that contains all the labels, where each label is repeated a number of times. Hence, the training symbol matrix can be represented as $\mathbf{X}_\text{t} = [\check{\mathbf{X}}_1, \check{\mathbf{X}}_2, \ldots , \check{\mathbf{X}}_K]$, 
where $\check{\mathbf{X}}_k = [\check{\mathbf{x}}_k,\ldots,\check{\mathbf{x}}_k] \in \mathbb{C}^{N_\text{t} \times L_\text{t}}$ consists of $L_\text{t}$ labels $\check{\mathbf{x}}_k$, $k \in \mathcal{K}$. Using this training method, the representative vector $\check{\mathbf{y}}_k$ can be learned empirically as
\begin{equation}
\check{\mathbf{y}}_k = \frac{1}{L_\text{t}}\sum_{t=1}^{L_\text{t}}\mathbf{y}[(k-1)L_\text{t}+t],
\label{eq_emprical_representative_vectors}
\end{equation}
where $\mathbf{Y}_\text{t} = \big[\mathbf{y}[1],\ldots,\mathbf{y}[T_\text{t}]\big]= Q_b(\mathbf{HX}_\text{t}+\mathbf{Z}_\text{t})$. The length of the training sequence is $T_\text{t} = KL_\text{t}$. This training method has been employed in \cite{jeon2017blind,Ly2018learning}.
\subsection{Subspace Training Method}
It is worth noting that the training sequence does not need to cover all the labels for the receiver to learn $\check{\mathcal{Y}}$ when the constellation $\mathcal{M}$ satisfies either of the following two conditions:
\begin{itemize}
	\item \textit{Condition} 1: $-x \in \mathcal{M}$, $\forall x \in \mathcal{M}$.
	\item \textit{Condition} 2: $\alpha x \in \mathcal{M}$, $\forall x \in \mathcal{M} \text{ and } \forall \alpha \in \{-1,j,-j\} $.
\end{itemize}
Although Condition 2 implies Condition 1 when $\alpha = -1$, i.e., any $\mathcal{M}$ satisfying Condition 2 will also satisfy Condition 1, we maintain these as two separate conditions for convenience in our later derivations. Examples of $\mathcal{M}$ for Condition 1 are BPSK, $8$-QAM and for Condition 2 are QPSK, $8$-PSK, $16$-QAM. 

If Condition 1 is satisfied, $- \check{\mathbf{x}}_k \in \check{\mathcal{X}}$ for all $\check{\mathbf{x}}_k \in \check{\mathcal{X}}$. The set of all labels can be written as
\begin{equation}
\check{\mathcal{X}} = \{\check{\mathcal{X}}_\text{ha},-\check{\mathcal{X}}_\text{ha}\},
\label{eq_cond1_training_seq}
\end{equation}
where $\check{\mathcal{X}}_\text{ha} = \{\check{\mathbf{x}}_1,\ldots,\check{\mathbf{x}}_{K/2}\}$. Without loss of generality, it is assumed that $\check{\mathbf{x}}_{k+K/2} = -\check{\mathbf{x}}_k$ with $k \in \{1,\ldots,K/2\}$.
If Condition 2 is satisfied, then $\alpha \check{\mathbf{x}}_k \in \check{\mathcal{X}}$ for all $ \check{\mathbf{x}}_k \in \check{\mathcal{X}}$ and $\alpha \in \{-1,j,-j\}$. The set of all labels can be written as
\begin{equation}
\begin{split}
\check{\mathcal{X}} = \{\check{\mathcal{X}}_\text{fo},-\check{\mathcal{X}}_\text{fo},j\check{\mathcal{X}}_\text{fo},-j\check{\mathcal{X}}_\text{fo}\},
\end{split}
\label{eq_cond2_training_seq}
\end{equation}
where $\check{\mathcal{X}}_\text{fo} = \{\check{\mathbf{x}}_1,\ldots,\check{\mathbf{x}}_{K/4}\}$. It is then assumed that $\check{\mathbf{x}}_{k+K/4} = -\check{\mathbf{x}}_k$, $\check{\mathbf{x}}_{k+K/2} = j\check{\mathbf{x}}_k$, and $\check{\mathbf{x}}_{k+3K/4} = -j\check{\mathbf{x}}_k$ for $k \in \{1,\ldots,K/4\}$.

The work in \cite{Jeon2018supervised} showed that if the transmitter employs QAM modulation and the quantization function satisfies $Q_b(-r) = - Q_b(r)$ $\forall r \in \mathbb{R}$, then the length of the training sequence can be reduced to $T_\text{t} = KL_\text{t}/4$. In Proposition \ref{theorem1} below, we generalize this result for any modulation scheme.
\begin{proposition}
	Given any constellation $\mathcal{M}$, if the quantizer $Q_b(.)$ is symmetric, i.e., $Q_b(-r) = - Q_b(r)$ $\forall r \in \mathbb{R}$, the length of the training sequence $T_\textup{t}$ can be reduced to
	\begin{equation}
	T_\textup{t} = \begin{cases}
	\frac{1}{2}KL_\textup{t} & \textup{if Condition 1 holds},\\
	\frac{1}{4}KL_\textup{t} & \textup{if Condition 2 holds}.
	\end{cases}
	\end{equation}
	\label{theorem1}
\end{proposition}
\begin{IEEEproof}
	Given any two labels $\check{\mathbf{x}}_{k_1}$ and $\check{\mathbf{x}}_{k_2} = -\check{\mathbf{x}}_{k_1}$, we have
	\begin{align}
	p(\mathbf{y}|\mathbf{x} = \check{\mathbf{x}}_{k_2}) &= \mathbb{P}[\mathbf{y} = Q_b(\mathbf{Hx}_{k_2}+\mathbf{z})]\nonumber \\
	&= \mathbb{P}[\mathbf{y} = Q_b(-\mathbf{Hx}_{k_1}-\mathbf{z})]\nonumber \\
	&= \mathbb{P}[-\mathbf{y} = Q_b(\mathbf{Hx}_{k_1}+\mathbf{z})]\nonumber \\
	&= p(-\mathbf{y}|\mathbf{x} = \check{\mathbf{x}}_{k_1}).
	\end{align}
	Therefore, $\check{\mathbf{y}}_{k_2} = - \check{\mathbf{y}}_{k_1}$ since
	\begin{align}
	\check{\mathbf{y}}_{k_2} = \mathbb{E}\big[\mathbf{y}|\mathbf{x} = \mathbf{\check{x}}_{k_2}\big] &= \sum \mathbf{y} p(\mathbf{y}|\mathbf{x} = \check{\mathbf{x}}_{k_2})\nonumber \\
	& = \sum \mathbf{y} p(-\mathbf{y}|\mathbf{x} = \check{\mathbf{x}}_{k_1})\nonumber \\
	& = -\sum \dot{\mathbf{y}} p(\dot{\mathbf{y}}|\mathbf{x} = \check{\mathbf{x}}_{k_1})\label{eq_change_var}\\
	& = -\mathbb{E}\big[\mathbf{y}|\mathbf{x} = \mathbf{\check{x}}_{k_1}\big] = - \check{\mathbf{y}}_{k_1}, \label{eq_same_space}
	\end{align}
	where (\ref{eq_change_var}) is obtained by setting $\dot{\mathbf{y}} = -\mathbf{y}$ and (\ref{eq_same_space}) holds because the sample spaces of $\dot{\mathbf{y}}$ and $\mathbf{y}$ are the same. Hence, the representative vectors satisfy  $\check{\mathbf{y}}_{k+K/2} = -\check{\mathbf{y}}_k$ with $k \in \{1,\ldots,K/2\}$ if Condition 1 holds. This means the training sequence only needs to cover $\check{\mathcal{X}}_\text{ha}$ to help the receiver learn all $K$ representative vectors in $\check{\mathcal{Y}}$. Similarly, when Condition 2 holds, we can also show that $\check{\mathbf{y}}_{k+K/4} = -\check{\mathbf{y}}_k$, $\check{\mathbf{y}}_{k+K/2} = j\check{\mathbf{y}}_k$, and $\check{\mathbf{y}}_{k+3K/4} = -j\check{\mathbf{y}}_k$ with $k \in \{1,\ldots,K/4\}$, and so the training sequence only needs to contain $\check{\mathcal{X}}_\text{fo}$. It should be noted that the proof for Condition 2 requires that $Q_b(jc) = jQ_b(c)$ $\forall c \in \mathbb{C}$, which is satisfied for our assumed quantizer.
\end{IEEEproof}
\section{Proposed Learning Methods}
The MCD detection method is simple but it has a primary drawback -- its detection accuracy heavily depends on the length of the training sequence. If the training sequence cannot provide accurate representative vectors in (\ref{eq_emprical_representative_vectors}), then detection errors will appear in (\ref{eq_MCD}). In fact, a short training sequence often results in poor estimation of the representative vectors. In order to improve the detection accuracy \textit{without} lengthening the training sequence, our idea is to use the training sequence as an initial guide for the learning process, and then find more precise representative vectors by exploiting other information.
\subsection{Proposed Supervised Learning Method}\label{AA}
\begin{figure}
	\centering
	\includegraphics[scale=0.39]{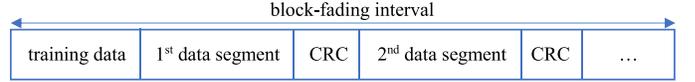}
	\caption{Usage of CRC for multiple data segments in each block-fading interval.}
	\label{fig_CRC_in_block_fading}
\end{figure}
In practical communications systems, error control mechanisms such as the CRC can be used to determine whether a segment of data is correctly decoded or not. This approach has been exploited to mitigate the effect of imperfect CSI on the ML detection for low-resolution ADCs \cite{jeon2018reinforcement}, \cite{jeon2018robust}. An error correcting code was also used to update the weights in a neural network as the channel changes, assuming perfect ADCs \cite{schibisch2018online}.

In our proposed method, should the CRC be available, it can be exploited for blind detection as follows: Data detection is first performed by the MCD using the training sequence, then the correctly decoded data confirmed by the CRC is used to augment the training set. As a result, the representative vectors obtained from the training sequence in (\ref{eq_emprical_representative_vectors}) can be refined and the incorrectly decoded data can be  re-evaluated by the MCD data detection. The process of CRC checking, updating the representative vectors, and data detection is repeated until no further correctly decoded segment is found.

In the system considered, we assume the use of the CRC for multiple data segments as illustrated in Fig. \ref{fig_CRC_in_block_fading}. Suppose there are $S$ segments in one block-fading interval, and each segment contains a data segment and a CRC block. Let $L_{\text{CRC}}$ and $L_\text{data}$ denote the length of the CRC and the length of each data segment in bits, respectively. Thus, we have
\begin{equation}
S\times(L_\text{data} + L_{\text{CRC}}) = T_\text{d} \times N_\text{t} \times \text{log}_2M.
\end{equation}
We also assume that $L_\text{data} + L_{\text{CRC}}$ is a multiple of $N_\text{t} \text{log}_2M$. This means the number of bits in a segment is a multiple of the number bits in a transmitted vector. The decoding algorithm of this proposed method is presented in Algorithm \ref{algo_supervised}. The detailed explanation of Algorithm \ref{algo_supervised} is as follows.

Let $\mathcal{Y}_k$ denote the set of received vectors that is used to estimate $\check{\mathbf{y}}_k$. Initially, $\mathcal{Y}_k$ is simply the set of received training vectors, i.e., $\mathcal{Y}_k = \big\{\mathbf{y}[(k-1)L_\text{t}+1],\ldots,\mathbf{y}[kL_\text{t}]\big\}$. Using these received training vectors, the representative vectors can be initially estimated (line \ref{algo1_inital_representative_vectors}). Let $s$ denote the index of the segments, $s \in\{1,2,\ldots,S\}$, and $\mathbf{Y}_s$ the $s^{\text{th}}$ received data segment. MCD detection is first performed on the given segment $\mathbf{Y}_s$ (line 6). If the CRC confirms the correct detection of $\mathbf{Y}_s$, the received vectors in $\mathbf{Y}_s$ are used to enlarge $\{\mathcal{Y}_k\}$. For example, if $\hat{k} = f(\mathbf{y}[n])$ is the decoded index of received vector $\mathbf{y}[n]$ (line \ref{algo1_detection_res}), then we can put $\mathbf{y}[n]$ into $\mathcal{Y}_{\hat{k}}$ (line \ref{algo1_enlarge_Dk}). In addition, based on the proof of Proposition \ref{theorem1}, we also use $\mathbf{y}[n]$ to augment other sets (line \ref{algo1_update_cond1} or \ref{algo1_update_cond2}) as follows:
\begin{itemize}
	\item \textit{Condition} 1:
	\begin{equation}
	\begin{cases}
	&\mathcal{Y}_{\hat{k}+\frac{K}{2}} = \mathcal{Y}_{\hat{k}+\frac{K}{2}} \cup \{-\mathbf{y}[n]\} \quad \text{if } \hat{k} \leq \frac{K}{2}, \\ 
	&\mathcal{Y}_{\hat{k}-\frac{K}{2}} = \mathcal{Y}_{\hat{k}-\frac{K}{2}} \cup \{-\mathbf{y}[n]\} \quad \text{if } \hat{k} > \frac{K}{2}.
	\end{cases}
	\label{eq_update_cond1}
	\end{equation}
	\item \textit{Condition} 2:\\
	Let $\mathcal{K}_1 = \{1,\ldots,\frac{K}{4}\}$, $\mathcal{K}_2 = \{\frac{K}{4}+1,\ldots,\frac{K}{2}\}$,\\$\mathcal{K}_3 = \{\frac{K}{2}+1,\ldots,\frac{3K}{4}\}$, and $\mathcal{K}_4 = \{\frac{3K}{4} + 1,\ldots,K\}$.\\
	If $\hat{k} \in \mathcal{K}_1$, let $\bar{k}_1 = \hat{k} + \frac{K}{4}$, $\bar{k}_2 = \hat{k} + \frac{K}{2}$, $\bar{k}_3 = \hat{k} + \frac{3K}{4}$.\\
	If $\hat{k} \in \mathcal{K}_2$, let $\bar{k}_1 = \hat{k} - \frac{K}{4}$, $\bar{k}_2 = \hat{k} + \frac{K}{2}$, $\bar{k}_3 = \hat{k} + \frac{K}{4}$.\\
	If $\hat{k} \in \mathcal{K}_3$, let $\bar{k}_1 = \hat{k} + \frac{K}{4}$, $\bar{k}_2 = \hat{k} - \frac{K}{4}$, $\bar{k}_3 = \hat{k} - \frac{K}{2}$.\\
	If $\hat{k} \in \mathcal{K}_4$, let $\bar{k}_1 = \hat{k} - \frac{K}{4}$, $\bar{k}_2 = \hat{k} - \frac{3K}{4}$, $\bar{k}_3 = \hat{k} - \frac{K}{2}$.\\
	Then three other sets can be updated as
	\begin{equation}
	\begin{cases}
	&\mathcal{Y}_{\bar{k}_1} = \mathcal{Y}_{\bar{k}_1} \cup \{-\mathbf{y}[n]\}, \\ 
	&\mathcal{Y}_{\bar{k}_2} = \mathcal{Y}_{\bar{k}_2} \cup \{j\mathbf{y}[n]\},\\
	&\mathcal{Y}_{\bar{k}_3} = \mathcal{Y}_{\bar{k}_3} \cup \{-j\mathbf{y}[n]\}.
	\end{cases}
	\label{eq_update_cond2}
	\end{equation}
\end{itemize}
After the training sets are enlarged, the representative vectors can be refined. Suppose $\mathcal{Y}_k = \{\mathbf{y}[n_{k,1}],\ldots,\mathbf{y}[n_{k,L_k}]\}$, the representative vector $\check{\mathbf{y}}_k$ is refined as
\begin{equation}
\check{\mathbf{y}}_k = \frac{1}{L_k}\sum_{t=1}^{L_k}\mathbf{y}[n_{k,t}],
\label{eq_update_representaive_vectors_supervised}
\end{equation}
with $n_{k,t} \in \{1,\ldots,T_\text{b}\}$ (line \ref{algo1_update_representaive_vectors_supervised}). It should be noted that $L_k \geq L_\text{t}$ $\forall k \in \mathcal{K}$. The refined representative vectors are then used to perform data detection on the next segment (back to line \ref{algo1_MCD}). In the first iteration, the next segment is $\mathbf{Y}_{s+1}$, which has not been decoded before. In the subsequent iterations, the next segment is one that has already been decoded incorrectly. Iterations here are accounted for by the \textbf{while} loop. The process of CRC checking, updating the representative vectors and data detection is repeated until no further correctly decoded segment is found or the received CRCs indicates that the whole received data block is decoded correctly (line \ref{algo1_stopping}).

Since this proposed method requires the use of the CRC, it can only be applied in systems where the CRC is available. In the following section, we propose another method, which does not require the use of the CRC, but can still obtain more precise input-output relations.
\begin{algorithm}[t]
	\small
	Initialize $\mathcal{C} = \varnothing$, $\mathcal{S} = \{1,2,\ldots,S\}$ and $stopping = false$\;
	Find $\check{\mathcal{Y}}$ using the training sequence \label{algo1_inital_representative_vectors}\;
	\While{stopping = false}{
		\For{$i = 1:length(\mathcal{S})$}{
			Let $s = \mathcal{S}(i)$\;
			Perform MCD detection on $\mathbf{Y}_s$ using $\check{\mathcal{Y}}$ \label{algo1_MCD}\;
			\If{\textup{CRC confirms the correct detection of} $\mathbf{Y}_s$}{
				Set $\mathcal{C} = \mathcal{C} \cup \{s\}$\;
				\ForEach{$\mathbf{y}[n] \in \mathbf{Y}_s$}{
					Let $\hat{k} = f(\mathbf{y}[n])$ \label{algo1_detection_res}\;
					Set $\mathcal{Y}_{\hat{k}} = \mathcal{Y}_{\hat{k}} \cup \{\mathbf{y}[n]\}$\label{algo1_enlarge_Dk}\;
					\If{\textup{Condition 1 holds}}{
						Perform (\ref{eq_update_cond1})\label{algo1_update_cond1}\;
					}
					\If{\textup{Condition 2 holds}}{
						Perform (\ref{eq_update_cond2})\label{algo1_update_cond2}\;
					}
				}
				Update $\check{\mathcal{Y}}$ using (\ref{eq_update_representaive_vectors_supervised})\label{algo1_update_representaive_vectors_supervised}\;
			}
		}
		\eIf{$\mathcal{C} = \varnothing$}{
			$stopping = true$ \label{algo1_stopping}\;
		}{
			Set $\mathcal{S} = \mathcal{S} \backslash \mathcal{C}$, then set $\mathcal{C} = \varnothing$\;
		}
	}
	\caption{Supervised Learning Decoding.}
	\label{algo_supervised}
\end{algorithm}
\subsection{Proposed Semi-supervised Learning Method}
In this part we propose a semi-supervised learning method. This proposed method is based on the K-means clustering technique \cite{bishop2006pattern}. The idea is to use the training sequence as an initial guidance to find coarse estimates of the representative vectors. Based on these coarse estimates, the received data vectors are then self-classified iteratively.

The K-means clustering technique aims to partition data into a number of clusters. However, in this communication context, the decoding task is not just to partition the received data into clusters but also to assign labels to the clusters, which can be done by using the training sequence. In addition, we take into account the constraints $\check{\mathbf{y}}_{k+K/2} = -\check{\mathbf{y}}_k$, $k = 1,\ldots,K/2$, if Condition 1 holds; and the constraints $\check{\mathbf{y}}_{k+K/4} = -\check{\mathbf{y}}_k$, $\check{\mathbf{y}}_{k+K/2} = j\check{\mathbf{y}}_k$, $\check{\mathbf{y}}_{k+3K/4} = -j\check{\mathbf{y}}_k$, $k = 1,\ldots,K/4$, if Condition 2 holds. These constraints can be adopted because clusters are formed based on their centroids, which are also referred to as the representative vectors $\{\check{\mathbf{y}}_k \}$ in this paper.

First, we introduce a set of binary variables $\beta_{n,k} \in \{0,1\}$ to indicate which of the $K$ labels that the received vector $\mathbf{y}[n]$ belongs to. Specifically, if a received vector $\mathbf{y}[n]$ belongs to label $k$, then $\beta_{n,k} = 1$ and $\beta_{n,l} = 0$ $\forall l \neq k$. We have the following optimization problems:
\begin{itemize}
	\item \textit{Condition 1}: 
	\begin{equation}
	\begin{aligned}
	& \underset{\{\beta_{n,k}\},\{\check{\mathbf{y}}_k\}}{\operatorname{minimize}} 
	& & J = \sum_{n=1}^{T_\text{b}} \sum_{k=1}^{K} \beta_{n,k} \|\mathbf{y}[n] - \check{\mathbf{y}}_k\|^2\\
	& \text{subject to}
	& & \check{\mathbf{y}}_{k+\frac{K}{2}} = - \check{\mathbf{y}}_k, \quad k = 1,\ldots,K/2.
	\end{aligned}
	\label{eq_opt_prob_cond1}
	\end{equation}
	The objective function in (\ref{eq_opt_prob_cond1}) is called the \textit{distortion measure} \cite{bishop2006pattern}. This optimization problem can be rewritten as
	\begin{equation}
	\begin{aligned}
	\underset{\{\beta_{n,k}\},\{\check{\mathbf{y}}_k\}}{\operatorname{minimize}} & & J_1
	\end{aligned}
	\label{eq_opt_prob_cond1_rewritten}
	\end{equation}
	where
	\begin{equation}
	J_1 = \sum_{n=1}^{T_\text{b}} \sum_{k=1}^{\frac{K}{2}} \big(\beta_{n,k} \|\mathbf{y}[n] - \check{\mathbf{y}}_k\|^2 +  \beta_{n,k+\frac{K}{2}}\|\mathbf{y}[n] + \check{\mathbf{y}}_k\|^2\big).
	\end{equation}
	Problem (\ref{eq_opt_prob_cond1_rewritten}) can be solved iteratively in which each iteration finds $\{\beta_{n,k}\}$ based on fixed $\{\check{\mathbf{y}}_k\}$ and vice versa. If $\{\check{\mathbf{y}}_k\}$ are fixed, $J_1$ is a linear function of $\{\beta_{n,k}\}$. It can be seen that the solutions $\{\beta_{n,k}\}$ are independent of $n$, so they can be found separately. For a given $n \in \{T_\text{t}+1,\ldots,T_\text{b}\}$, the optimization problem for $\{\beta_{n,k}\}$ is 
	\begin{equation}
	\begin{aligned}
	& \underset{\{\beta_{n,k}\}}{\operatorname{minimize}} 
	& & \sum_{k=1}^{K} \beta_{n,k} \|\mathbf{y}[n] - \check{\mathbf{y}}_k\|^2,
	\end{aligned}
	\end{equation}
	whose solution is found by setting $\beta_{n,k} = 1$ for the $k$ associated with the minimum value of $\|\mathbf{y}[n] - \check{\mathbf{y}}_k\|^2$. The solutions $\{\beta_{n,k}\}$ can be written as 
	\begin{equation}
	\beta_{n,k} = \begin{cases}
	1 & \text{if } k = \operatorname{argmin}_{k'} \|\mathbf{y}[n] - \check{\mathbf{y}}_{k'}\|^2,\\
	0 & \text{otherwise}.
	\end{cases}
	\label{eq_gamma_sol}
	\end{equation}
	It should be noted that $\beta_{n,k} = 1$ whenever $n \leq T_{\text{t}}$ and $k = \lfloor (n-1)/L_\text{t} \rfloor+1$ because the labels of the received training vectors are known at the receiver.
	When the $\{\beta_{n,k}\}$ are fixed, $J_1$ becomes a quadratic function of $\{\check{\mathbf{y}}_k\}$. Hence the solutions $\{\check{\mathbf{y}}_k\}$ can be found by finding the derivative of $J_1$ with respect to $\check{\mathbf{y}}_k$:
	\begin{equation}
	\frac{\partial J_1}{\partial \check{\mathbf{y}}_k} = \sum_{n=1}^{T_\text{b}} \beta_{n,k}\big(-\mathbf{y}[n]^H+ \check{\mathbf{y}}_k^H\big) + \beta_{n,k+\frac{K}{2}}\big(\mathbf{y}[n]^H + \check{\mathbf{y}}_k^H\big),
	\end{equation}
	when being set to $0$ yields
	\begin{equation}
	\check{\mathbf{y}}_k = \frac{\sum_n \big(\beta_{n,k} - \beta_{n,k+\frac{K}{2}}\big)\mathbf{y}[n]}{\sum_n \big(\beta_{n,k} + \beta_{n,k+\frac{K}{2}}\big)}, \quad k = 1,\ldots,\frac{K}{2}.
	\label{eq_centroid_cond1}
	\end{equation}
	Equation (\ref{eq_centroid_cond1}) says that the representative vector $\check{\mathbf{y}}_k$, with $k\leq K/2$, is calculated by using the received vectors that not only belong to cluster $k$ but also to cluster $k+K/2$.
	\item \textit{Condition 2}:
	\begin{equation}
	\begin{aligned}
	& \underset{\{\beta_{n,k}\},\{\check{\mathbf{y}}_k\}}{\operatorname{minimize}} 
	& & J = \sum_{n=1}^{T_\text{b}} \sum_{k=1}^{K} \beta_{n,k} \|\mathbf{y}[n] - \check{\mathbf{y}}_k\|^2\\
	& \text{subject to}
	& & \check{\mathbf{y}}_{k+\frac{K}{4}} = - \check{\mathbf{y}}_k\\
	& & & \check{\mathbf{y}}_{k+\frac{K}{2}} = j \check{\mathbf{y}}_k\\
	& & & \check{\mathbf{y}}_{k+\frac{3K}{4}} = -j \check{\mathbf{y}}_k\\
	& & & k = 1,\ldots,K/4.
	\end{aligned}
	\label{eq_opt_prob_cond2}
	\end{equation}
	The optimization problem (\ref{eq_opt_prob_cond2}) can also be rewritten as
	\begin{equation}
	\begin{aligned}
	\underset{\{\beta_{n,k}\},\{\check{\mathbf{y}}_k\}}{\operatorname{minimize}} & & J_2
	\end{aligned}
	\label{eq_opt_prob_cond2_rewritten}
	\end{equation}
	where
	\begin{equation}
	\begin{split}
	J_2 &= \sum_{n=1}^{T_\text{b}} \sum_{k=1}^{\frac{K}{4}} \big(\beta_{n,k} \|\mathbf{y}[n] - \check{\mathbf{y}}_k\|^2 + \beta_{n,k+\frac{K}{4}}\|\mathbf{y}[n] + \check{\mathbf{y}}_k\|^2 \\
	& + \beta_{n,k+\frac{K}{2}}\|\mathbf{y}[n] - j\check{\mathbf{y}}_k\|^2 + \beta_{n,k+\frac{3K}{4}}\|\mathbf{y}[n] + j\check{\mathbf{y}}_k\|^2\big)
	\end{split}
	\end{equation}
	Applying the same technique as in Condition 1 to this problem, we can find $\beta_{n,k}$ from (\ref{eq_gamma_sol}) and 
	\begin{equation}
	\begin{split}
	\check{\mathbf{y}}_k = &\frac{\sum_n \big(\beta_{n,k} - \beta_{n,k+\frac{K}{4}}-j\beta_{n,k+\frac{K}{2}}+j\beta_{n,k+\frac{3K}{4}}\big)\mathbf{y}[n]}{\sum_n \big(\beta_{n,k} + \beta_{n,k+\frac{K}{4}}+\beta_{n,k+\frac{K}{2}}+\beta_{n,k+\frac{3K}{4}}\big)},\\
	& \quad k = 1,\ldots,\frac{K}{4}.
	\end{split}
	\label{eq_centroid_cond2}
	\end{equation}
	Equation (\ref{eq_centroid_cond2}) also points out that the representative vector $\check{\mathbf{y}}_k$, with $k \leq K/4$, is found by using the received vectors that not only belong to cluster $k$ but also to clusters $k+K/4$, $k+K/2$ and $k+3K/4$.
\end{itemize}
\begin{algorithm}[t]
	\small
	Initialize $stopping = false$, $iter = 0$\;
	Find $\mathcal{Y}$ using the training sequence\label{algo2_coarse_centroids}\;
	\While{stopping = false}{
		$iter = iter + 1$\;
		Perform (\ref{eq_gamma_sol})\label{algo2_clustering}\;
		\If{\textup{Condition 1 holds}}{
			Perform (\ref{eq_centroid_cond1})\label{algo2_update1_cond1}\;
			Set $\check{\mathbf{y}}_{k+\frac{K}{2}} = - \check{\mathbf{y}}_k,$ with $k = 1,\ldots,K/2$\label{algo2_update2_cond1}\;
		}
		\If{\textup{Condition 2 holds}}{
			Perform (\ref{eq_centroid_cond2})\label{algo2_update1_cond2}\;
			Set $\check{\mathbf{y}}_{k+\frac{K}{4}} = - \check{\mathbf{y}}_k$, $\check{\mathbf{y}}_{k+\frac{K}{2}} = j \check{\mathbf{y}}_k$, $\check{\mathbf{y}}_{k+\frac{3K}{4}} = -j \check{\mathbf{y}}_k$,
			with $k = 1,\ldots,K/4$\label{algo2_update2_cond2}\;
		}
		\If{\textup{convergent or} $iter = iter_\textup{max}$}{$stopping = true$ \label{algo2_stopping}\;}
	}
	\caption{Semi-supervised Learning Decoding.}
	\label{algo_semisupervised}
\end{algorithm}

The decoding algorithm for this semi-supervised learning method is presented in Algorithm \ref{algo_semisupervised}. Coarse estimation of the representative vectors is first obtained by using the training sequence (line \ref{algo2_coarse_centroids}). Then clustering is applied on all of the received data vectors (line \ref{algo2_clustering}). Depending on whether Condition 1 or Condition 2 is satisfied, the representative vectors are updated (lines \ref{algo2_update1_cond1}-\ref{algo2_update2_cond1} or lines \ref{algo2_update1_cond2}-\ref{algo2_update2_cond2}). The process of clustering the received data vectors and updating the representative vectors is repeated until convergence or the number of iterations exceeds a maximum value (line \ref{algo2_stopping}). Convergence is achieved if the solutions $\{\beta_{n,k}\}$ are the same for two successive iterations. Convergence of algorithm 2 is assured because after each iteration, the value of the objective function does not increase. However, the point of convergence is not guaranteed to be a global optimum.
\section{Performance Analysis with One-bit ADCs}
\label{sec_performance_analysis}
This section presents a performance analysis of the proposed methods for the case of $1$-bit ADCs. We assume that all symbol vectors in $\mathcal{X}$ are a priori equally likely to be transmitted.
The objective is to characterize the VER. Since the performance of our proposed methods for $1$-bit ADCs is independent of the step size $\Delta$, we choose $\Delta = 2$ so that the quantization function becomes the $\operatorname{sign}(.)$ function, where $\operatorname{sign}(a)=+1$ if $a \geq 0$ and $\operatorname{sign}(a)=-1$ if $a < 0$. If $a$ is a complex number, then $\operatorname{sign}(a)$ = $\operatorname{sign}(\Re\{a\}) + j\operatorname{sign}(\Im\{a\})$. The operator $\operatorname{sign}(.)$ of a matrix or vector is applied separately  to every element of that matrix or vector.
\vspace{-0.2cm}
\subsection{VER Analysis at Low SNRs}
Here, we present an approximate pairwise VER at low SNRs for the Rayleigh fading channel. First, using the Bussgang decomposition, the system model $\mathbf{y} = Q_b(\mathbf{r})$ can be rewritten as $\mathbf{y} = \mathbf{F}\mathbf{r} + \mathbf{e}$ \cite{mezghani2012capacity}
where $\mathbf{e}$ is the quantization distortion and 
\begin{equation}
\mathbf{F} = \sqrt{\frac{2}{\pi}}\operatorname{diag}(\boldsymbol{\Sigma}_{r})^{-\frac{1}{2}}.
\end{equation}
The term $\mathbf{\Sigma}_{r} = \mathbf{H}\mathbf{H}^H + N_0\mathbf{I}$ is the covariance matrix of $\mathbf{r}$.
Let $\mathbf{A} = \mathbf{F}\mathbf{H}$ and $\mathbf{w} = \mathbf{F}\mathbf{z} + \mathbf{e}$, then the system model becomes
\begin{equation}
\mathbf{y} = \mathbf{A}\mathbf{x} + \mathbf{w},
\end{equation}
where $\mathbf{A} = \sqrt{2/\pi}\operatorname{diag}(\boldsymbol{\Sigma}_{r})^{-\frac{1}{2}}\mathbf{H}$
and the effective noise $\mathbf{w} = [w_1,w_2,\ldots,w_{N_\text{r}}]^T$ is modeled as Gaussian \cite{mezghani2012capacity} with zero mean and covariance matrix
\begin{equation}
\begin{split}
\!\!\!\!\!\!\boldsymbol{\Sigma}_{w} = &\frac{2}{\pi}\Big[\operatorname{arcsin}\Big(\operatorname{diag}(\mathbf{\Sigma}_{r})^{-\frac{1}{2}}\mathbf{\Sigma}_{r}\operatorname{diag}(\mathbf{\Sigma}_{r})^{-\frac{1}{2}}\Big)-\\
&\quad \operatorname{diag}(\mathbf{\Sigma}_{r})^{-\frac{1}{2}}\mathbf{\Sigma}_{r}\operatorname{diag}(\mathbf{\Sigma}_{r})^{-\frac{1}{2}}\!+\!N_0\operatorname{diag}(\mathbf{\Sigma}_{r})^{-1}\Big].\!\!\!\!
\end{split}
\end{equation}
Note that the operation $\operatorname{arcsin}(\mathbf{.})$ of a matrix is applied element-wise on that matrix. The representative vector $\check{\mathbf{y}}_k$ now becomes $\check{\mathbf{y}}_k = \mathbf{A}\check{\mathbf{x}}_k$.

In the low SNR regime, the approximation $\mathbf{\Sigma}_{r} \approx \mathbf{\Sigma}_{z}$ holds \cite{mezghani2012capacity}, where $\mathbf{\Sigma}_{z} = N_0\mathbf{I}$ is the covariance matrix of $\mathbf{z}$. This approximation leads to $\mathbf{A} \approx \sqrt{2/(N_0\pi)}\mathbf{H}$ and $\boldsymbol{\Sigma}_{w} \approx \mathbf{I}$. 
Let $\boldsymbol{\upsilon} = [\upsilon_1,\ldots,\upsilon_{N_\text{r}}]^T = \check{\mathbf{y}}_k-\check{\mathbf{y}}_{k'}$, where $\upsilon_{i} = \sqrt{2/(N_0\pi)}\mathbf{h}^T_i(\check{\mathbf{x}}_k - \check{\mathbf{x}}_{k'})$ with $\mathbf{h}_i$ being the $i^{\text{th}}$ column of $\mathbf{H}$. Since $\mathbf{H}$ is comprised of i.i.d. Gaussian random variables $\mathcal{CN}(0,1)$, $\upsilon_{i}$ is also Gaussian of zero mean with variance
\begin{equation}
\sigma^2_{kk'} = \frac{2}{N_0\pi}\|\check{\mathbf{x}}_{k} - \check{\mathbf{x}}_{k'}\|_2^2. \label{distance}
\end{equation}

Denote $P_{\check{\mathbf{x}}_k\rightarrow \check{\mathbf{x}}_{k'}}$ as the pairwise vector error probability of confusing $\check{\mathbf{x}}_k$ with $\check{\mathbf{x}}_{k'}$ when $\check{\mathbf{x}}_k$ is transmitted and when  $\check{\mathbf{x}}_{k}$ and $\check{\mathbf{x}}_{k'}$ are the only two hypotheses \cite{tse2005fundamentals}. The following proposition establishes the relationship between $P_{\check{\mathbf{x}}_k\rightarrow \check{\mathbf{x}}_{k'}}$  and $\sigma^2_{kk'}$.
\begin{proposition}\label{proposition-0}
$P_{\check{\mathbf{x}}_k\rightarrow \check{\mathbf{x}}_{k'}}$ at low SNR can be approximated as
\begin{eqnarray}
P_{\check{\mathbf{x}}_k\rightarrow \check{\mathbf{x}}_{k'}} &\approx  1- \Phi \Big(\sqrt{N_\mathrm{r}/(1 +  2/\sigma^2_{kk'})}\Big). 
\label{eq_pairwise_VER_low_SNRs}
\end{eqnarray}
\end{proposition}
\begin{IEEEproof}
Please refer to Appendix \ref{append-A}.
\end{IEEEproof}
The result in Proposition \ref{proposition-0} clearly shows the dependency of the pairwise VER  on the Euclidean distance between the two symbol vectors $\check{\mathbf{x}}_k$ and $\check{\mathbf{x}}_{k'}$. We now proceed to obtain an upper bound on the VER, denoted as $P^{\text{ver}}_{\rho}$, at low SNR assuming a priori equally likely $\check{\mathbf{x}}_1,\ldots,\check{\mathbf{x}}_K$.
\begin{proposition}\label{proposition-1}
$P^{\mathrm{ver}}_{\rho}$ at low SNR is upper-bounded as
\begin{eqnarray}\label{VER-upper}
P^{\mathrm{ver}}_{\rho} &\leq& \frac{1}{K}\sum_{k=1}^{K}  \sum_{k'\neq k}^K\left[1- \Phi \Big(\sqrt{N_\text{r}/(1 +  2/\sigma^2_{kk'})}\Big)\right].
\end{eqnarray}
\end{proposition}
\begin{IEEEproof}
The bound on $P^{\mathrm{ver}}_{\rho}$ is obtained via the union bound
\begin{eqnarray}
\!\!\!P^{\text{ver}}_{\rho} = \sum_{k=1}^{K}\mathbb{P}[\hat{\mathbf{x}}\neq  \check{\mathbf{x}}_k, \mathbf{x} = \check{\mathbf{x}}_k]
&=&\frac{1}{K}\sum_{k=1}^{K}\mathbb{P}[\hat{\mathbf{x}}\neq  \check{\mathbf{x}}_k \bigm\vert \mathbf{x}= \check{\mathbf{x}}_k]\nonumber \\
&\leq& \frac{1}{K}\sum_{k=1}^{K} \sum_{k'\neq k}^K P_{\check{\mathbf{x}}_k\rightarrow \check{\mathbf{x}}_{k'}} \nonumber
\end{eqnarray}
and the application of Proposition \ref{proposition-0}.
\end{IEEEproof}


The probability $\mathbb{P}[\hat{\mathbf{x}}\neq \check{\mathbf{x}}_k \,|\,\mathbf{x} = \check{\mathbf{x}}_k]$ is invariant to $\check{\mathbf{x}}_k$ for the case of PSK modulation. Without loss of generality, we assume that $\check{\mathbf{x}}_1$ was transmitted, so that the VER simplifies to  
\begin{equation}
P^{\text{ver}}_{\rho} \leq \sum_{k\neq 1}^K\left[1- \Phi \Big(\sqrt{N_\text{r}/(1 +  2/\sigma^2_{1k})}\Big)\right].
\end{equation}
We note that this result is valid for low SNRs. In the following analysis, we characterize the VER at a very high SNR, i.e., $\rho \rightarrow \infty$.

\subsection{VER Analysis as $\rho \rightarrow \infty$}
Here we evaluate the VER as the SNR tends to infinity. Let $\mathbf{g}_k = [g_{k,1},\ldots, g_{k,N_\text{r}}]^T = \mathbf{H}\check{\mathbf{x}}_k$, then
\begin{align}
\mathbb{P}[\Re\{y_i\}=+1\mid \mathbf{x} = \check{\mathbf{x}}_k] &= \Phi(\sqrt{2\rho/N_\text{t}}\,\Re\{g_{k,i}\}), \label{eq_rx_sym_prob1}\\
\mathbb{P}[\Im\{y_i\}=+1\mid \mathbf{x} = \check{\mathbf{x}}_k] &= \Phi(\sqrt{2\rho/N_\text{t}}\,\Im\{g_{k,i}\}). \label{eq_rx_sym_prob2}
\end{align}
The true representative vectors are
\begin{equation}
\check{\mathbf{y}}_{k} = \mathbb{E}\big[\mathbf{y}\mid\mathbf{x} = \check{\mathbf{x}}_k\big] = 2\Phi (\sqrt{2\rho/N_\text{t}}\mathbf{g}_k) - (\mathbf{1}+j\mathbf{1})
\end{equation}
which becomes $\mathrm{sign}(\mathbf{g}_k)$ as $\rho\rightarrow \infty$.
It is possible for a given realization of $\mathbf{H}$ that more than one symbol vector will lead to the same representative vector: $\mathrm{sign}(\mathbf{g}_k) = \mathrm{sign}(\mathbf{g}_{k'})$ with $k \neq k'$, and in such cases a detection error will occur regardless of the detection scheme. In the following, we analyze the probability that $\mathrm{sign}(\mathbf{g}_k) = \mathrm{sign}(\mathbf{g}_{k'})$. Our analysis is applicable for the cases of BPSK and QPSK modulation. 

To facilitate the analysis, we convert the notation into the real domain as follows:
\begin{align*}
&\check{\mathbf{x}}_k^{\Re} = [\check{x}_{k,1}^{\Re},\check{x}_{k,2}^{\Re},\ldots,\check{x}_{k,2N_\text{t}}^{\Re}]^T = [\Re \{\check{\mathbf{x}}_k\}^T, \Im \{\check{\mathbf{x}}_k\}^T]^T,\\
&\mathbf{g}_k^{\Re} = [g_{k,1}^{\Re},g_{k,2}^{\Re},\ldots,g_{k,2N_\text{r}}^{\Re}]^T = [\Re \{\mathbf{g}_k\}^T, \Im \{\mathbf{g}_k\}^T]^T.
\end{align*}

We first consider BPSK modulation, i.e., $\mathcal{M} = \{\pm 1\}$. In this case, $\Im \{\check{\mathbf{x}}_k\} = \mathbf{0}$. 
\begin{theorem}\label{theorem-2}
Given $d = \|\check{\mathbf{x}}_k^{\Re}-\check{\mathbf{x}}_{k'}^{\Re}\|_0$ as the Hamming distance between the two labels,  then
\begin{equation}
\mathbb{P}\big[\mathrm{sign}(\mathbf{g}_k) = \mathrm{sign}(\mathbf{g}_{k'})\big] = \Bigg[\frac{2}{\pi}\operatorname{arctan}\sqrt{\frac{N_{\mathrm{t}}-d}{d}}\Bigg]^{2N_\text{r}}.
\label{eq_g_1_equals_g_2}
\end{equation}
\end{theorem}
\begin{IEEEproof}
Please refer to Appendix \ref{append-C}.
\end{IEEEproof}

As $\rho\rightarrow \infty$, the effect of the AWGN can be ignored. Thus, $\mathbb{P}\big[\check{\mathbf{y}}_{k} = \check{\mathbf{y}}_{k'}\big]=   \mathbb{P}\big[\mathrm{sign}(\mathbf{g}_k) = \mathrm{sign}(\mathbf{g}_{k'})\big]$. An upper bound on the VER is established in the following proposition.
\begin{proposition}\label{proposition-2}
With BPSK modulation, the asymptotic VER at high SNR is upper-bounded as
\begin{eqnarray}\label{upper-bound}
P^{\mathrm{ver}}_{\rho \rightarrow \infty} \leq \frac{1}{2} \sum_{d=1}^{N_{\mathrm{t}}} \binom{N_{\mathrm{t}}}{d}\Bigg[\frac{2}{\pi}\operatorname{arctan}\sqrt{\frac{N_{\mathrm{t}}-d}{d}}\Bigg]^{2N_\text{r}}.
\end{eqnarray}
\end{proposition}
\begin{IEEEproof}
Please refer to Appendix \ref{append-D}.
\end{IEEEproof}

\begin{proposition}\label{proposition-3}
With BPSK modulation and $N_{\mathrm{t}} = 2$, the upper bound in \eqref{upper-bound} is tight.
\end{proposition}
\begin{IEEEproof}
For BPSK modulation and $N_{\mathrm{t}} = 2$, let $\check{\mathbf{x}}_1^{\Re} = [1,1,0,0]$, $\check{\mathbf{x}}_2^{\Re} = [1,-1,0,0]$, $\check{\mathbf{x}}_3^{\Re} = [-1,1,0,0]$, $\check{\mathbf{x}}_4^{\Re} = [-1,-1,0,0]$. Herein, $\check{\mathbf{x}}_1^{\Re} = -\check{\mathbf{x}}_4^{\Re}$ and $\check{\mathbf{x}}_2^{\Re} = -\check{\mathbf{x}}_3^{\Re}$, resulting in
$\check{\mathbf{y}}_1 = -\check{\mathbf{y}}_4$ and $\check{\mathbf{y}}_2 = -\check{\mathbf{y}}_3$ as $\rho\rightarrow \infty$. Hence, events $\check{\mathbf{y}}_1 = \check{\mathbf{y}}_{2}$
and $\check{\mathbf{y}}_1 = \check{\mathbf{y}}_{3}$ are mutually exclusive while event $\check{\mathbf{y}}_1 = \check{\mathbf{y}}_{4}$ does not exist. This proposition thus follows as a direct consequence of the proof for Proposition \ref{proposition-2} given in Appendix \ref{append-D}.
\end{IEEEproof}

For the case of QPSK modulation, the Hamming distance $d = \|\check{\mathbf{x}}_k^{\Re}-\check{\mathbf{x}}_{k'}^{\Re}\|_0$ between any two labels can be as large as $2N_{\mathrm{t}}$. Following the same derivation as in Theorem \ref{theorem-2} and Proposition \ref{proposition-2}, an upper-bound for the asymptotic VER at high SNR can be established by the following proposition.
\begin{proposition}\label{proposition-4}
With QPSK modulation, the asymptotic VER at high SNR is upper-bounded as
\begin{eqnarray}
P^{\mathrm{ver}}_{\rho \rightarrow \infty} \leq \frac{1}{2} \sum_{d=1}^{2N_{\mathrm{t}}} \binom{2N_{\mathrm{t}}}{d}\Bigg[\frac{2}{\pi}\operatorname{arctan}\sqrt{\frac{2N_{\mathrm{t}}-d}{d}}\Bigg]^{2N_\text{r}}.
\end{eqnarray}
\end{proposition}

\subsection{Transmit Signal Design}
\label{sec_tx_signal_design}
Thus far it has been assumed that the transmitter uses all $K$ possible labels for transmission. However, as $K$ grows large, the training task for all the $K$ labels becomes impractical, since the block fading interval $T_{\text{b}}$ is finite. 
In this section, we consider a system where the transmitter employs only a subset of $\tilde{K}$ labels among the $K$ possible labels for both the training and data transmission phases. The rest of the $K - \tilde{K}$ labels are unused. 
While using only $\tilde{K}$ labels reduces the transmission rate as compared to using all the $K$ possible labels, the VER can be improved. In many 5G networks, e.g., Machine-to-Machine (M2M) communication systems, the priority is on the reliability, not the rate \cite{Boccardi2014Five}. 
In addition, the reduction in training time with small $\tilde{K}$ may help improve the system throughput.

The design problem is how to choose $\tilde{K}$ labels among the $K$ labels. To address this problem, let us look back at Proposition \ref{proposition-2} and Proposition \ref{proposition-4}. These propositions reveal that the VER at infinite SNR is inversely proportional to the Hamming distances between the labels. Therefore, the $\tilde{K}$ labels should be chosen such that their Hamming distances are as large as possible. Based on this observation, we propose the following criterion for choosing the transmit signals:
\begin{equation}
\mathcal{X}^{\star} = \argmax_{\mathcal{X} \subset \check{\mathcal{X}}^{\Re}} \underset{1\leq k_1 < k_2 \leq \tilde{K}}{\operatorname{min}} \|\mathbf{x}_{k_1} - \mathbf{x}_{k_2}\|_0,
\label{eq_tx_criterion}
\end{equation}
where $\mathcal{X} = \{\mathbf{x}_1,\ldots,\mathbf{x}_{\tilde{K}}\}$ denote the set of $\tilde{K}$ different labels for transmission, and $\check{\mathcal{X}}^{\Re} = \{\check{\mathbf{x}}_1^{\Re},\ldots,\check{\mathbf{x}}_{K}^{\Re}\}$. 
This design criterion aims to maximize the minimum Hamming distance. Note that the proposed criterion is also applicable for low SNRs because as shown in Proposition \ref{proposition-1}, the VER is inversely proportional to the Euclidean distance, which is analogous to the Hamming distance for BPSK and QPSK, albeit with some scaling factor. It should be noted that the proposed criterion does not rely on a specific channel realization, so the design task can be carried out off-line. 

Problem (\ref{eq_tx_criterion}) can be solved by exhaustive search when $\binom{K}{\tilde{K}}$ is not too large. When the exhaustive search is not possible, we propose a simple greedy algorithm, whose pseudo-code can be found in Algorithm \ref{algo_tx_signal_design}. Here, $d_{\text{min}}(\mathcal{X})$ denotes the objective function of (\ref{eq_tx_criterion}) and $\mathcal{X}'$ in line \ref{algo3_unused_label} denotes the set of labels, which is not used for transmission. The principle of Algorithm \ref{algo_tx_signal_design} is as follows: 
\begin{itemize}
	\item Generate $N$ initial sets $\{\mathcal{X}_i\}_{i = 1,\ldots,N}$, where each set $\mathcal{X}_i$ contains $\tilde{K}$ different labels randomly chosen from $\check{\mathcal{X}}^{\Re}$.
	\item For each initial set $\mathcal{X}_i$, find $\mathbf{x}' \in \mathcal{X}'$ such that when an element of $\mathcal{X}_i$ is replaced by
	$\mathbf{x}'$, the objective function value is increased. This is repeated until no further increase in the objective function is possible after evaluating all replacements.
	\item Each initial set $\mathcal{X}_i$ produces a corresponding solution $\mathcal{X}^{\star}_i$ as in line \ref{algo3_solution_Xi}. The solution $\mathcal{X}^{\star}$ of (\ref{eq_tx_criterion}) is obtained by selecting the $\mathcal{X}^{\star}_i$ whose objective function value is largest (line \ref{algo3_solution_X}).
\end{itemize}
Note that the larger $N$ is, the more likely Algorithm \ref{algo_tx_signal_design} will find the optimal solution.
\begin{algorithm}[t]
	\small
	Randomly generate $N$ initial sets $\{\mathcal{X}_i, i = 1, \ldots, N\}$\;
	\For{$i=1:N$}{
		$stopping = false$\;
		\While{stopping = false}{
			Let $flag = 1$\;
			Set $\mathcal{X}' = \check{\mathcal{X}} \backslash \mathcal{X}_i = \{\mathbf{x}'_1,\ldots,\mathbf{x}'_{K - \tilde{K}}\}$ \label{algo3_unused_label}\;
			\For{$k_1 = 1:\tilde{K}$} {
				\For{$k_2 = 1:K-\tilde{K}$} {
					Let $\hat{\mathcal{X}}_i = \big(\mathcal{X}_i \backslash \{\mathbf{x}_{k_1}\}\big)\cup \big\{\mathbf{x}'_{k_2}\big\}$\;
					\If{ $d_{\textup{min}}(\hat{\mathcal{X}}_i) > d_{\textup{min}}(\mathcal{X}_i)$\label{algo3_check_increase}}{
						Set $\mathcal{X}_i = \hat{\mathcal{X}}$ and $flag = 0$\;
						Exit both \textbf{for} loops\;
					}
				}
			}
			\If{$flag = 1$}{
				Set $stopping = true$ and $\mathcal{X}^{\star}_i = \mathcal{X}_i$\label{algo3_stopping} \label{algo3_solution_Xi}\;
			}
		}
	}
	$\mathcal{X}^{\star}=\argmax_{\mathcal{X}^{\star}_i} d_{\text{min}}(\mathcal{X}^{\star}_i)$ \label{algo3_solution_X}\;
	\caption{Transmit Signal Design.}
	\label{algo_tx_signal_design}
\end{algorithm}
\section{Simulations and Results}
\subsection{Numerical Evaluation of the Proposed Methods}
\label{sec_sim_and_results}
\begin{figure*}[t!]
	\centering
	\begin{subfigure}[t]{0.5\textwidth}
		\centering
		\includegraphics[scale=0.63]{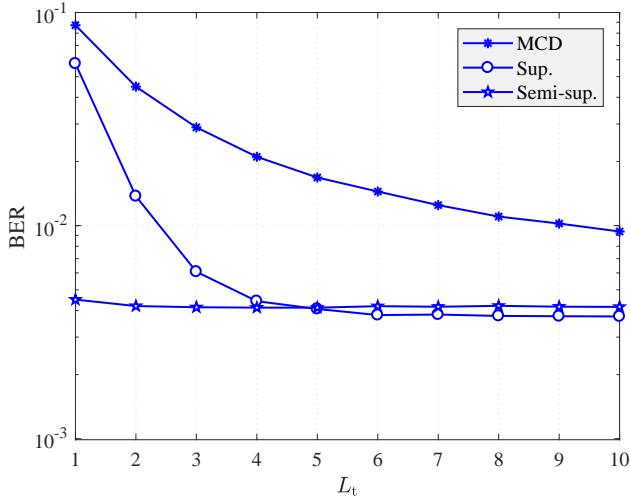}
		\caption{$L_\text{t}$ varies and $\rho = 0$ dB.}
		\label{fig_effect_of_Lt_a}
	\end{subfigure}%
	\begin{subfigure}[t]{0.5\textwidth}
		\centering
		\includegraphics[scale=0.63]{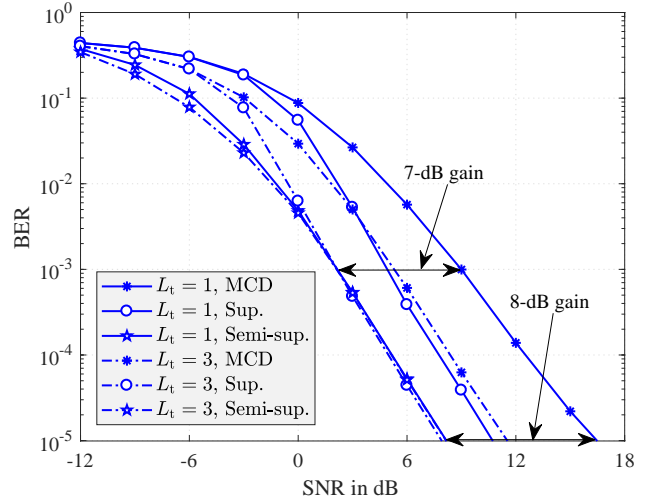}
		\caption{$L_\text{t}=1$ and $L_\text{t}=3$, $\rho$ varies. }
		\label{fig_effect_of_Lt_b}
	\end{subfigure}
	\caption{Effect of $L_\text{t}$ on MCD and the proposed methods with 1-bit ADCs, $N_\text{r} = 16$ and BPSK modulation.}
	\label{fig_effect_of_Lt}
\end{figure*}
\begin{figure*}[t!]
	\centering
	\begin{subfigure}[t]{0.5\textwidth}
		\centering
		\includegraphics[scale=0.63]{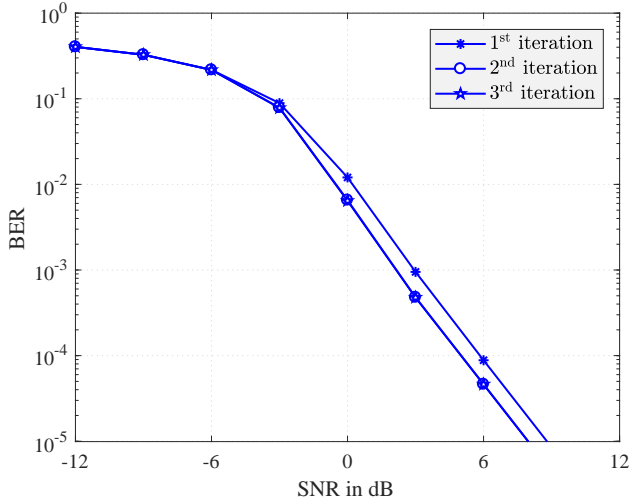}
		\caption{Supervised learning method.}
		\label{fig_supervised_improvement_over_iteration}
	\end{subfigure}%
	\begin{subfigure}[t]{0.5\textwidth}
		\centering
		\includegraphics[scale=0.63]{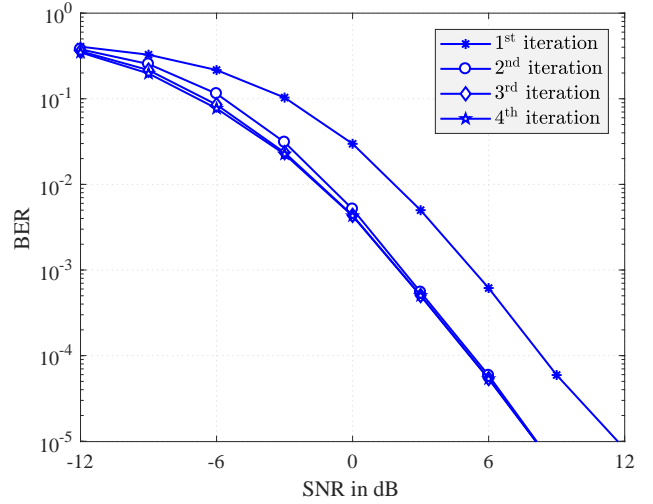}
		\caption{Semi-supervised learning method.}
		\label{fig_semisupervised_improvement_over_iteration}
	\end{subfigure}
	\caption{Performance improvement for different iterations with $1$-bit ADCs, BPSK modulation, $N_\text{r}=16$ and $L_\text{t} = 3$.}
	\label{fig_improvement_over_iteration}
\end{figure*}
We use Monte Carlo simulations to numerically evaluate the performance of our proposed methods. The simulation settings are as follows. The number of transmit antennas $N_\text{t}$ is set to be $2$ unless otherwise stated. The data phase contains $T_\text{d} = 500$ time slots. In the supervised learning method, we adopt a $24$-bit CRC as in the $3$GPP Long Term Evolution (LTE) standard \cite{3GPP_standard}. The generator of the CRC in our simulation is $z^{24} + z^{23} + z^{14} + z^{12} + z^8 + 1$, and the length of each data segment is $16$ bits. Thus, the length of each coded segment is $40$ bits. This is the minimum length in the $3$GPP LTE standard. In all figures, `Sup.' and `Semi-sup.' stand for the supervised learning and semi-supervised learning methods, respectively.

We first study the effect of the training sequence length $L_\text{t}$ on MCD and the two proposed methods (Fig. \ref{fig_effect_of_Lt}). We use BPSK modulation with $N_\text{r} = 16$ and 1-bit ADCs. Fig. \ref{fig_effect_of_Lt_a} shows the change of the BER as $L_\text{t}$ varies. An interesting observation is that the performance of the semi-supervised learning method is much less affected by $L_\text{t}$ compared to the two other methods. Hence, the length of the training sequence can be reduced without causing much degradation on the performance of the semi-supervised learning method. This is illustrated more clearly in Fig. \ref{fig_effect_of_Lt_b}, where we carry out the simulation for $L_\text{t} = 1$ and $L_\text{t} = 3$, still with BPSK modulation, 1-bit ADCs and $N_\text{r} = 16$. It can be seen from Fig. \ref{fig_effect_of_Lt_b} that, as $L_\text{t}$ is reduced from $3$ to $1$, the BER of MCD is significantly degraded while the BER of the semi-supervised learning method does not change for SNRs greater than $0$ dB and only minor degradation occurs for SNRs less than $0$ dB. This leads to a significant improvement for the semi-supervised learning method as compared to MCD, for example, about a $7$-dB gain at a BER of $10^{-3}$ and $8$-dB at a BER of $10^{-5}$ when $L_{\text{t}} = 1$. Even for moderately long training sequences, e.g., $L_{\text{t}} = 3$, the gain of our proposed methods is still considerable, from $3$-dB to $4$-dB.

\begin{figure*}
	\centering
	\begin{subfigure}[t]{0.5\textwidth}
		\centering
		\includegraphics[scale=0.63]{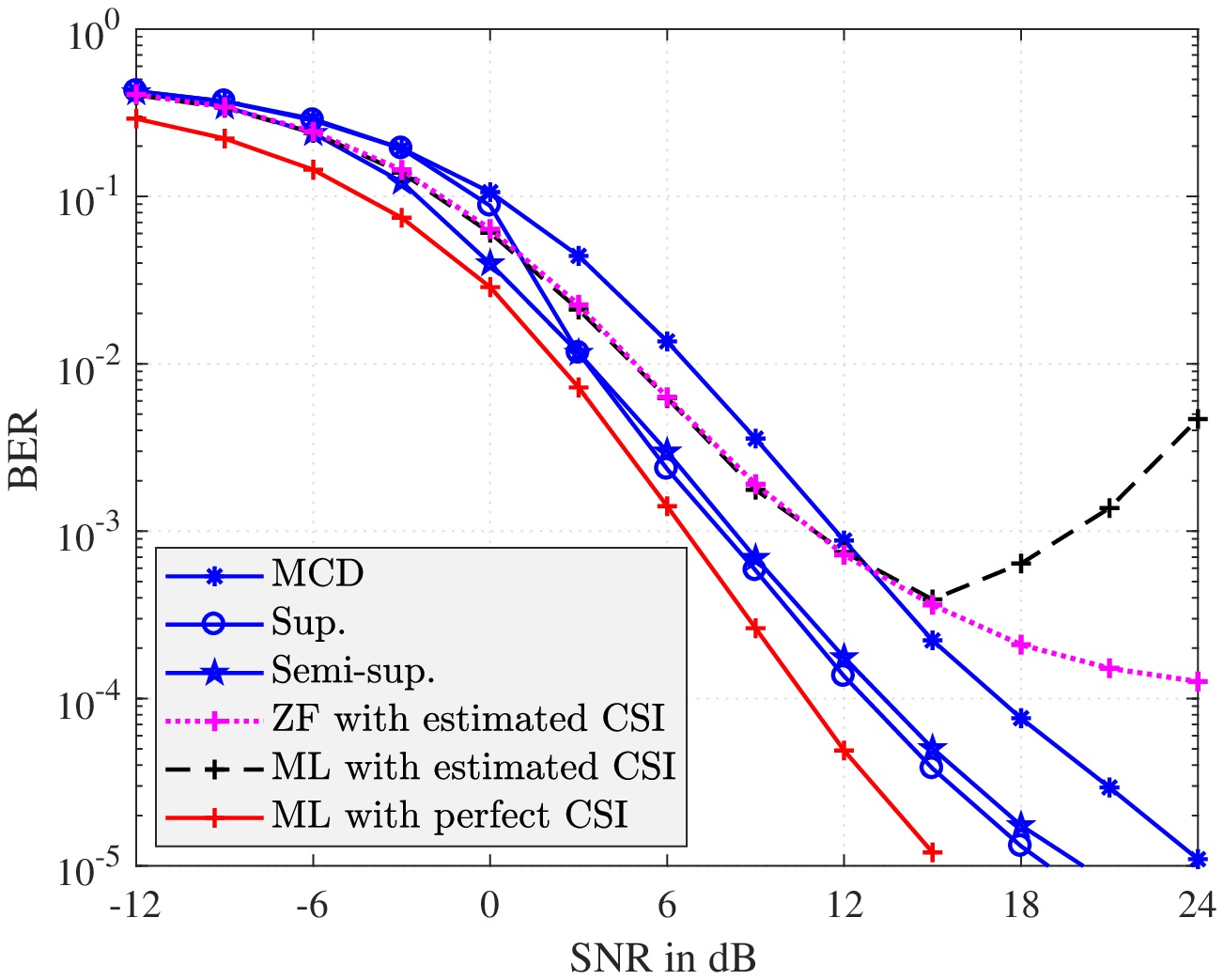}
		\caption{BER}
		\label{fig_BER_comparision_with_coherent_detection}
	\end{subfigure}%
	\begin{subfigure}[t]{0.5\textwidth}
		\centering
		\includegraphics[scale=0.63]{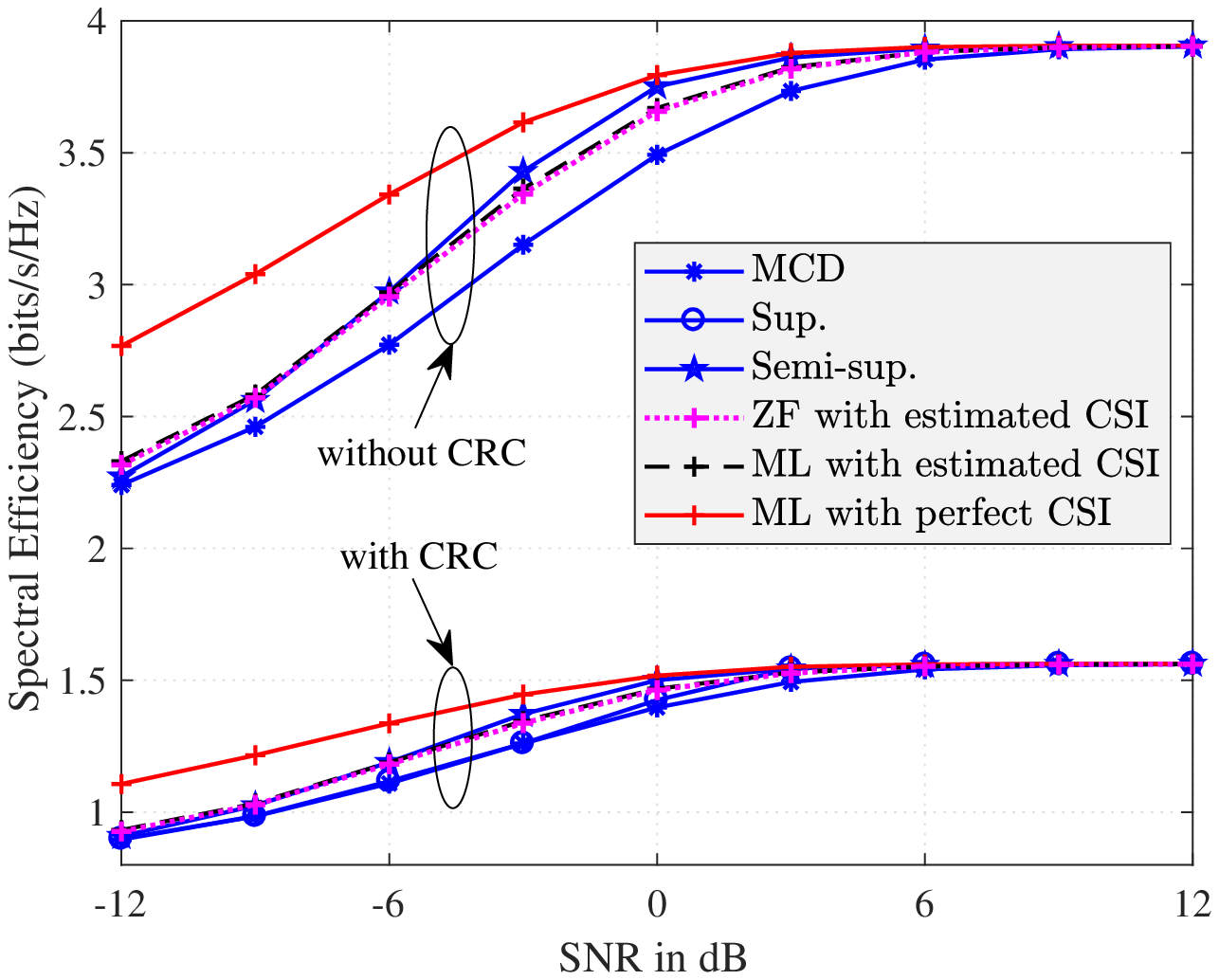}
		\caption{Spectral efficiency}
		\label{fig_rate_comparision_with_coherent_detection}
	\end{subfigure}
	\caption{Performance comparison between blind and coherent detection with $1$-bit ADCs, QPSK modulation, $N_\text{r}=16$ and $L_\text{t} = 3$.}
	\label{fig_comparision_with_coherent_detection}
\end{figure*}
\begin{figure*}
	\centering
	\begin{subfigure}[t]{0.5\textwidth}
		\centering
		\includegraphics[scale=0.63]{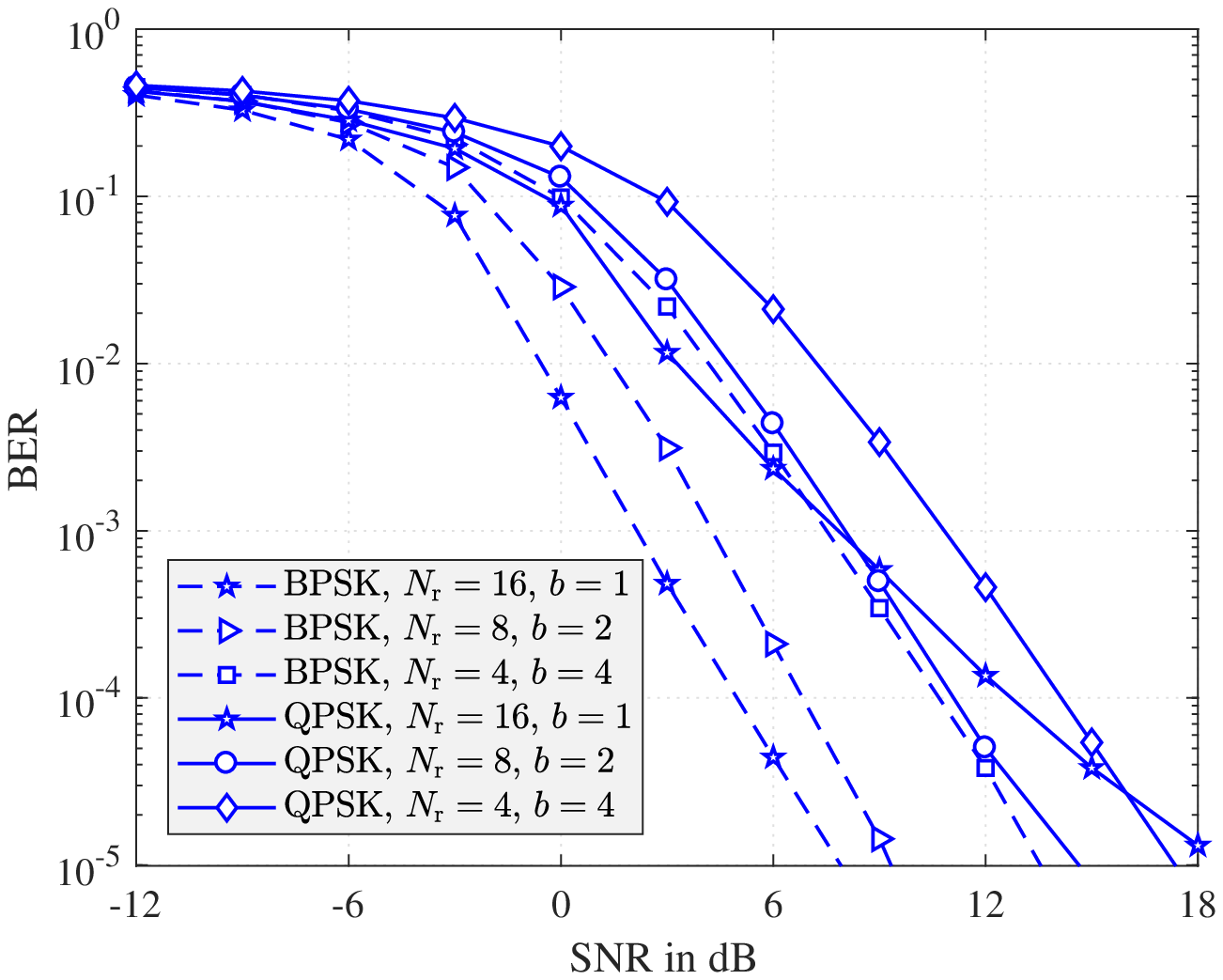}
		\caption{Supervised learning method.}
		\label{fig_supervised_Nr_vs_b_tradeoff}
	\end{subfigure}%
	\begin{subfigure}[t]{0.5\textwidth}
		\centering
		\includegraphics[scale=0.63]{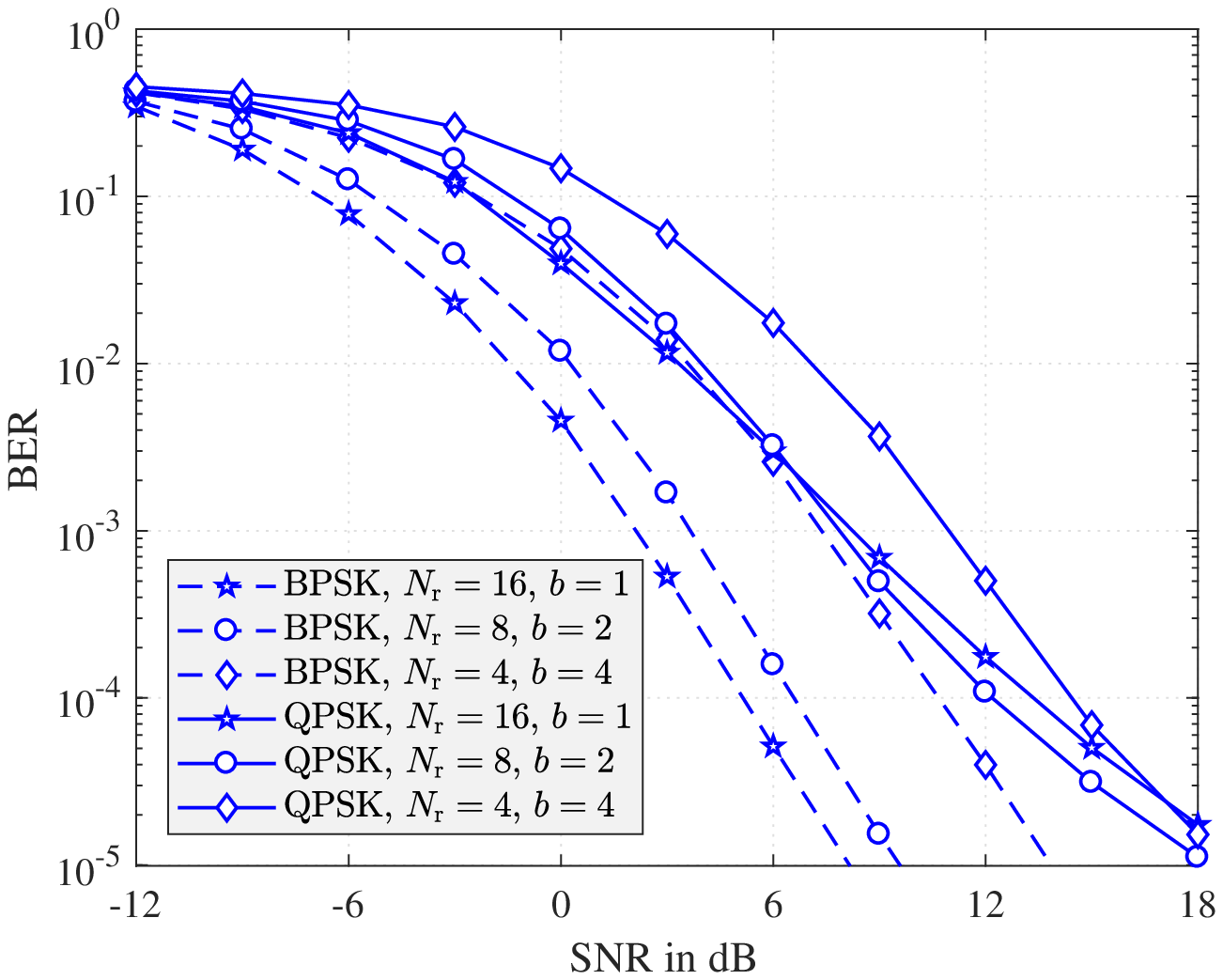}
		\caption{Semi-supervised learning method.}
		\label{fig_semi_supervised_Nr_vs_b_tradeoff}
	\end{subfigure}
	\caption{Performance of the proposed methods for different numbers of receive antennas $N_\text{r}$ and ADC resolutions $b$ with $L_\text{t} = 3$.}
	\label{fig_Nr_vs_b_tradeoff}
\end{figure*}

The results in Fig. \ref{fig_effect_of_Lt} can be explained as follows. The performance of MCD is susceptible to $L_\text{t}$ because its detection accuracy relies on the representative vectors estimated only from the training sequence. Therefore, if $L_\text{t}$ is small, the representative vectors are not estimated correctly and so the performance can be degraded significantly. On the other hand, the semi-supervised learning method is much less dependent on $L_\text{t}$ because it uses the training sequence only as the initial guide for the detection task. The representative vectors can still be refined using the to-be-decoded data. Compared to the semi-supervised learning method, the supervised learning method is more dependent on $L_\text{t}$ because as mentioned earlier, it depends on detection results from the training sequence.

Since the proposed methods work iteratively, we perform simulations to evaluate the improvement in BER over the iterations. Simulation results are shown in Fig. \ref{fig_improvement_over_iteration}. For the supervised learning method, Fig. \ref{fig_supervised_improvement_over_iteration}, it can be seen that the BER converges after only $2$ iterations. For the semi-supervised learning method, Fig. \ref{fig_semisupervised_improvement_over_iteration}, there is considerable improvement between the first and the second iterations, but then the third and the fourth iterations give approximately the same performance. It is therefore preferred to limit the maximum number of iterations to $3$ in the semi-supervised learning method. It should be noted that the BER on the first iteration of the semi-supervised learning method is actually the BER of the MCD method because the first iteration can only exploit the training sequence.

In Fig. \ref{fig_comparision_with_coherent_detection}, we compare the aforementioned blind detection methods with several coherent detection methods. The simulation uses $1$-bit ADCs, QPSK modulation, $N_\text{r}=16$ and $L_\text{t} = 3$. For coherent detection, CSI is first estimated by the Bussgang Linear Minimum Mean Squared Error (BLMMSE) method proposed in \cite{li2017channel}. The length of the training sequence in the blind detection methods is $12$, so we also set the length of the pilot sequence for the channel estimation to $12$. The ZF detection method is presented in \cite{li2017channel}. The ML method for $1$-bit ADCs is provided in \cite{Choi2015Quantized,choi2016near}. A performance comparison in terms of BER is given in Fig. \ref{fig_BER_comparision_with_coherent_detection}, which shows that the proposed methods outperform the ZF and ML methods with estimated CSI. It is also seen that the BER of our proposed methods is quite close the BER of ML detection with perfect CSI. Here, we observe a significant increase in the BER at high SNRs for the ML method with estimated CSI. This observation was also reported in \cite{jeon2017blind}. In comparing the two proposed methods in Fig. \ref{fig_BER_comparision_with_coherent_detection} and Fig. \ref{fig_effect_of_Lt}, should the CRC be available, it is more beneficial to use the supervised learning method for better BER performance.

Fig. \ref{fig_rate_comparision_with_coherent_detection} provides a comparison in terms of spectral efficiency $\eta$, defined as the average number of information bits received correctly per block-fading interval $T_{\mathrm{b}}$. We determine $\eta$ for the case without CRC as
\begin{eqnarray*}
\eta = \frac{T_{\mathrm{d}}}{T_{\mathrm{b}}}\times (1-\mathrm{BER})\times N_{\mathrm{t}}\times \log_2M
\end{eqnarray*}
and for the case with CRC as
\begin{eqnarray*}
\eta = \frac{L_{\mathrm{data}}}{L_{\mathrm{data}}+L_{\mathrm{CRC}}}\times \frac{T_{\mathrm{d}}}{T_{\mathrm{b}}}\times (1-\mathrm{BER})\times N_{\mathrm{t}}\times \log_2M
\end{eqnarray*}
Fig. \ref{fig_rate_comparision_with_coherent_detection} indicates a proportional drop in the spectral efficiency due to the use of CRC. Note that the supervised learning method can only be applied in systems where the CRC is available but the other methods can be used in any system regardless of the CRC. Thus, should the CRC be eliminated for improved spectral efficiency, the semi-supervised method provides better performance than MCD and conventional coherent detection with estimated CSI.


To study the trade-off between $N_\text{r}$ and $b$, we evaluate the proposed methods in three different scenarios: (i) $N_\text{r} = 4, b = 4$; (ii) $N_\text{r} = 8, b = 2$; and (iii) $N_\text{r} = 16, b = 1$. This is to ensure the same number of bits after the ADCs for baseband processing. The number of label repetitions $L_\text{t}$ is set to be $3$. The simulation results are shown in Fig. \ref{fig_Nr_vs_b_tradeoff}, with the supervised learning method in Fig. \ref{fig_supervised_Nr_vs_b_tradeoff} and the semi-supervised learning method in Fig. \ref{fig_semi_supervised_Nr_vs_b_tradeoff}. For BPSK modulation, the best performance is achieved by scenario (iii) for both methods. Hence, this suggests the use of more receive antennas and fewer bits in the ADCs when BPSK modulation is employed. However, for QPSK modulation, there is a trade-off between scenarios (ii) and (iii). For low SNRs, the setting $N_\text{r} = 16$ and $b = 1$ gives better performance, but for high SNRs, the best results are with $N_\text{r} = 8$ and $b = 2$.
\subsection{Validation of Performance Analysis}
\begin{figure}
	\centering
	\includegraphics[scale=0.6]{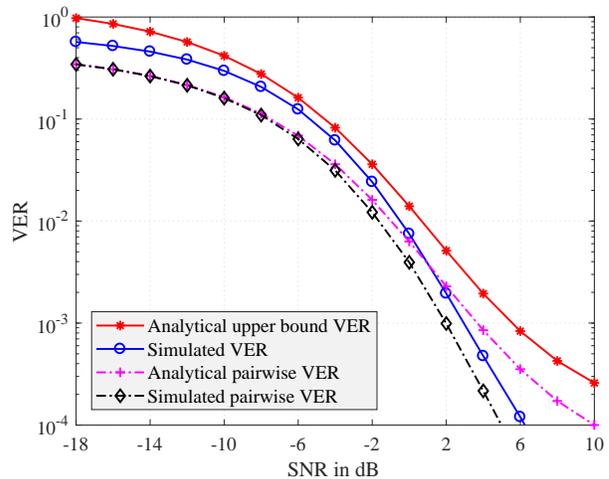}
	\caption{Validation of the analytical pairwise VER in (\ref{eq_pairwise_VER_low_SNRs}) and the analytical VER in (\ref{VER-upper}) at low SNRs with $N_\text{t}=2$, $N_\text{r} = 16$, and BPSK modulation.}
	\label{fig_VER_low_SNRs}
\end{figure}
\begin{figure}
	\centering
	\includegraphics[scale=0.6]{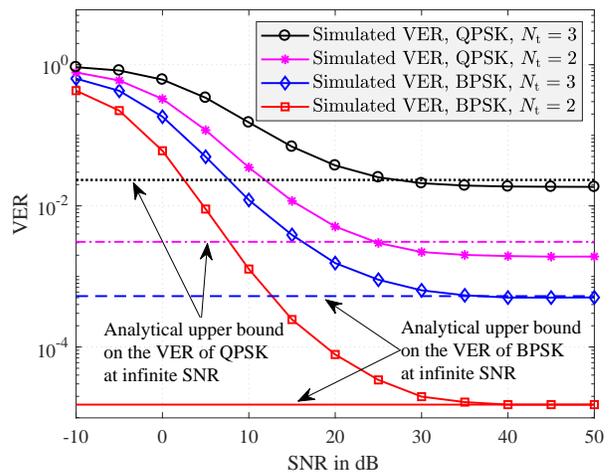}
	\caption{Validation of the analytical VER at infinite SNR in Propositions \ref{proposition-2}, \ref{proposition-3}, and \ref{proposition-4}.}
	\label{fig_validation_VER_infinite_SNR}
\end{figure}
\begin{figure}
	\centering
	\includegraphics[scale=0.6]{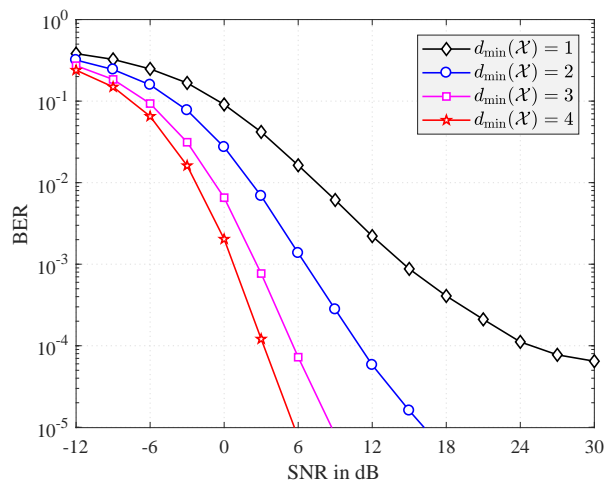}
	\caption{Validation of the transmit signal design with $N_\text{t} = 6$, $N_\text{r} = 16$, $\tilde{K} = 4$, and BPSK modulation.}
	\label{fig_tx_signal_design_result}
\end{figure}
This section presents a validation on the performance analyses in Section \ref{sec_performance_analysis}. Fig. \ref{fig_VER_low_SNRs} provides the analytical approximate pairwise VER in (\ref{eq_pairwise_VER_low_SNRs}) and the VER in (\ref{VER-upper}). We use the setting of $N_\text{t} = 2$, $N_\text{r} = 16$, and BPSK modulation. The two labels used to examine the pairwise VER are $\check{\mathbf{x}}_k = [+1, +1]^T$ and $\check{\mathbf{x}}_{k'} = [+1, -1]^T$. It can be seen that our approximate pairwise VER is very close to the simulated pairwise VER at low SNRs, typically with SNRs less than $0$-dB. However, as the SNR increases, our approximate pairwise VER tends to diverge from the true pairwise VER because the approximation $\boldsymbol{\Sigma}_r \approx \boldsymbol{\Sigma}_z$ is inapplicable for high SNRs. The simulation results also show that our analytical VER is quite close to the true VER at low SNRs.

Validation of the high SNR expressions for the analytical VER is given in Fig. \ref{fig_validation_VER_infinite_SNR} with $N_\text{r} = 8$. The horizontal lines represent the analytical upper bounds on the VER at infinite SNR. For the case of BPSK and $N_{\text{t}} = 2$, it can be seen that the simulated VER approaches the horizontal solid line as the SNR increases and then they match at very high SNRs. This validates the result of Proposition \ref{proposition-3} indicating that the bound is tight in the case of BPSK and $N_{\text{t}} = 2$. With BPSK and $N_{\text{t}} = 3$, the horizontal dashed line is just slightly higher than the floor of the simulated VER. For QPSK modulation, there is a small gap between the horizontal lines and the floors of the simulated VER. These observations validate our analytical upper-bound results in Proposition \ref{proposition-2} and Proposition \ref{proposition-4}.

In Fig. \ref{fig_tx_signal_design_result}, we provide a validation for the proposed transmit signal design. We assume perfectly learned representative vectors. Detection results are obtained by using MCD with these perfectly learned representative vectors. It can be seen that the BER is reduced as the minimum Hamming distance of $\mathcal{X}$ increases, which validates our analysis. In this particular simulation scenario ($N_\text{t} = 6$, $N_\text{r} = 16$, $\tilde{K} = 4$, and BPSK modulation), the minimum Hamming distance of an optimal set $\mathcal{X}^\star$ is $4$. Hence, the red star line also represents the BER obtained with an optimal $\mathcal{X}^\star$.

\begin{figure}
	\centering
	\includegraphics[scale=0.6]{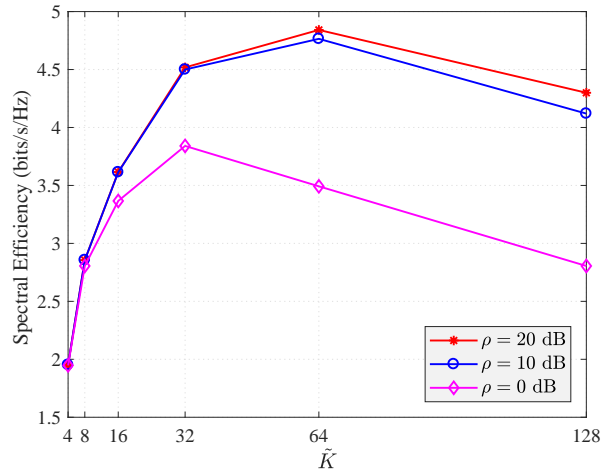}
	\caption{Spectral efficiency versus $\tilde{K}$ with $N_\text{t} = 8$, $N_\text{r} = 16$, BPSK modulation, $L_\text{t} = 3$, and $T_b = 500$.}
	\label{fig_spectgral_efficiency_vs_Ktilde}
\end{figure}

Finally, we examine the change of spectral efficiency with respect to $\tilde{K}$ at different SNR values. Simulations are carried out with $N_\text{t} = 8$, $N_\text{r} = 16$, BPSK modulation, $L_\text{t} = 3$, $T_b = 500$, and $\tilde{K}\in \{4,8,16,32,64,128\}$. Simulation results are given in Fig. \ref{fig_spectgral_efficiency_vs_Ktilde}, where the spectral efficiency is computed as
\begin{equation*}
	\eta = \frac{T_\text{d}} {T_\text{b}}\times(1-\mathrm{BER})\times \log_2\tilde{K}.
\end{equation*}

For each value of $\tilde{K}$, Algorithm \ref{algo_tx_signal_design} is applied to find the solution $\mathcal{X}^*$ of (\ref{eq_tx_criterion}). We found that, with $\tilde{K} \in \{4,8,16\}$, the symbol vectors of $\mathcal{X}^*$ do not satisfy condition 1, and so the full-space training method has to be used for these $\tilde{K}$ values. For  $\tilde{K} \in \{32,64,128\}$, the minimum Hamming distance of the optimal set is $d_\text{min}(\mathcal{X}^*) = 1$. Hence we can choose $\tilde{K}$ symbol vectors to meet condition 1 and so the subspace training method can be used to reduce the training overhead. We use the semi-supervised learning method for the detection task. The simulation results in Fig. \ref{fig_spectgral_efficiency_vs_Ktilde} show that increasing $\tilde{K}$ does not necessarily improve the spectral efficiency, due to the increased training overhead. There is thus an optimal value of $\tilde{K}$ that gives the highest spectral efficiency. For example, at  $0$-dB SNR, $\tilde{K} = 32$ is optimal, whereas for higher SNRs of $10$-dB or $20$-dB, $\tilde{K} = 64$ should be chosen.

\section{Conclusion}
In this paper, we have proposed two new learning methods for enhancing the performance of blind detection in MIMO systems that employ low-resolution ADCs. The supervised learning method exploits the use of CRC in practical systems to gain more training data. The semi-supervised learning method is based on the perspective that the to-be-decoded data can itself help the detection task thanks to grouping of received symbol vectors for the same transmitted signal. Simulation results demonstrate the performance improvement and robustness of our proposed methods over existing techniques. Numerical results also show that the two proposed learning methods require only a few iterations to converge. We have also carried out a performance analysis for the proposed methods by evaluating the VER in different SNR regimes. In addition, a new criterion for the transmit signal design problem has  also been proposed.

\appendices
\section{Proof of Proposition \ref{proposition-0}} \label{append-A}
We first express $P_{\check{\mathbf{x}}_k\rightarrow \check{\mathbf{x}}_{k'}}$ as follows:
\begin{align}
P_{\check{\mathbf{x}}_k\rightarrow \check{\mathbf{x}}_{k'}} &= \mathbb{P}\Big[\|\mathbf{y}-\check{\mathbf{y}}_{k}\|^2_2 \geq \|\mathbf{y}-\check{\mathbf{y}}_{k'}\|^2_2 \bigm\vert \mathbf{x} = \check{\mathbf{x}}_k\Big]\nonumber \\
&= \mathbb{P}\Big[\|\boldsymbol{\upsilon}\|^2_2 + 2\Re\{\boldsymbol{\upsilon}^H\mathbf{w}\} \leq 0\Big]\nonumber \\
&= \mathbb{P}\Big[\sum_{i=1}^{N_\text{r}}\big(|\upsilon_{i}|^2 + 2\Re \{\upsilon_{i}^*w_i\}\big)\leq 0\Big] \label{eq_37}.
\end{align}
By letting $\varepsilon_i = |\upsilon_{i}|^2 + 2\Re \{\upsilon_{i}^*w_i\}$, (\ref{eq_37}) becomes
\begin{equation}
P_{\check{\mathbf{x}}_k\rightarrow \check{\mathbf{x}}_{k'}} = \mathbb{P}\Big[\sum_{i=1}^{N_\text{r}}\varepsilon_i\leq 0\Big] \label{eq_39}.
\end{equation}
In order to approximate the probability in (\ref{eq_39}), we need to compute the mean and variance of $\varepsilon_i$. The mean of $\varepsilon_i$ is
\begin{align}
\mathbb{E}[\varepsilon_i] &= \mathbb{E}\big[|\upsilon_{i}|^2 + 2\Re \{\upsilon_{i}^*w_i\}\big] = \mathbb{E}\big[|\upsilon_{i}|^2 \big] = \sigma^2_{kk'}.
\end{align}
The variance of $\varepsilon_i$ is given as
\begin{equation}
\begin{split}
\sigma^2_{\varepsilon_i}&=\operatorname{Var}\big[|\upsilon_{i}|^2\big] + \operatorname{Var}\big[2\Re \{\upsilon_{i}^*w_i\}\big] +\\
&\phantom{{}=1}2\operatorname{Cov}\big(|\upsilon_{i}|^2,2\Re \{\upsilon_{i}^*w_i\}\big).
\end{split}
\label{eq_41}
\end{equation}
The first term in the right-hand side of (\ref{eq_41}) is
\begin{align}
\operatorname{Var}\big[|\upsilon_{i}|^2\big] &= \mathbb{E}\big[|\upsilon_{i}|^4\big] - \mathbb{E}\big[|\upsilon_{i}|^2\big]^2 = \sigma^4_{kk'} \label{eq_42}.
\end{align}
The second term in the right-hand side of (\ref{eq_41}) is
\begin{equation}
\begin{split}
\operatorname{Var}\big[2\Re \{\upsilon_{i}^*w_i\}\big] &= \operatorname{Var}\big[\upsilon_{i}^*w_i\big] + \operatorname{Var}\big[\upsilon_{i}w^*_i\big] +\\
&\phantom{{}=1}2\operatorname{Cov}\big(\upsilon_{i}^*w_i,\upsilon_{i}w^*_i\big).
\end{split}
\end{equation}
Since $\operatorname{Var}\big[\upsilon_{i}^*w_i\big] = \operatorname{Var}\big[\upsilon_{i}w^*_i\big] = \mathbb{E}\big[|\upsilon_{i}|^2 \big] = \sigma^2_{kk'}$, and $\operatorname{Cov}\big(\upsilon_{i}^*w_i,\upsilon_{i}w^*_i\big)  = 0$, we have
\begin{equation}
\operatorname{Var}\big[2\Re \{\upsilon_{i}^*w_i\}\big] = 2\sigma^2_{kk'}\label{eq_44}.
\end{equation} 
The last term in the right-hand side of (\ref{eq_41}) is 
\begin{equation}
\begin{split}
&\operatorname{Cov}\big(|\upsilon_{i}|^2,2\Re \{\upsilon_{i}^*w_i\}\big)=\\
&\phantom{{}=1}\mathbb{E}\big[|\upsilon_{i}|^22\Re \{\upsilon_{i}^*w_i\}\big] + \mathbb{E}\big[|\upsilon_{i}|^2\big]\mathbb{E}\big[2\Re \{\upsilon_{i}^*w_i\}\big]=0\label{eq_45},
\end{split}
\end{equation}
since $ \mathbb{E}\big[|\upsilon_{i}|^22\Re \{\upsilon_{i}^*w_i\}\big] = \mathbb{E}\big[|\upsilon_{i}|^2(\upsilon_{i}^*w_i+\upsilon_{i}w_i^*)\big]=0$ and $\mathbb{E}\big[2\Re \{\upsilon_{i}^*w_i\}\big] = \mathbb{E}\big[\upsilon_{i}^*w_i\big]+\mathbb{E}\big[\upsilon_{i}w_i^*\big] =0$.

Substituting the results in (\ref{eq_42}), (\ref{eq_44}), and (\ref{eq_45}) into (\ref{eq_41}) yields the variance of $\varepsilon_i$ as
\begin{equation}
\sigma^2_{\varepsilon_i} = \sigma^4_{kk'} +  2\sigma^2_{kk'}.
\end{equation}
The variables $\{\varepsilon_i\}_{i=1,\ldots,N_\text{r}}$ are i.i.d. because of the i.i.d. elements in $\mathbf{H}$. Hence, by the central limit theorem, the variable $\sum_{i=1}^{N_\text{r}}\varepsilon_i$ in (\ref{eq_39}) can be approximated by a Gaussian random variable with mean $N_\text{r}\sigma^2_{kk'}$ and variance $N_\text{r}(\sigma^4_{kk'} +  2\sigma^2_{kk'})$. Finally, the probability in (\ref{eq_39}) can be approximated as
\begin{align}
P_{\check{\mathbf{x}}_k\rightarrow \check{\mathbf{x}}_{k'}} &\approx \Phi \bigg(\frac{-N_\text{r}\sigma^2_{kk'}}{\sqrt{N_\text{r}(\sigma^4_{kk'} +  2\sigma^2_{kk'})}}\bigg)\nonumber \\
&= 1- \Phi \Big(\sqrt{N_\text{r}/(1 +  2/\sigma^2_{kk'})}\Big).
\end{align}

\section{Proof of Theorem \ref{theorem-2}}\label{append-C}
For two labels $\check{\mathbf{x}}^{\Re}_{k}$ and $\check{\mathbf{x}}^{\Re}_{k'}$ ,  we can always find two disjoint index sets $\mathcal{I}_{\mathrm{c}}$ and $\mathcal{I}_{\mathrm{d}}$ such that $\check{x}^{\Re}_{k,i} = \check{x}^{\Re}_{k',i} \neq 0$, $\forall i \in \mathcal{I}_{\mathrm{c}}$, and $\check{x}^{\Re}_{k,i} = -\check{x}^{\Re}_{k',i}$ $\forall i \in \mathcal{I}_{\mathrm{d}}$. We denote $d = |\mathcal{I}_{\mathrm{d}}|$ as the Hamming distance between the two labels $\check{\mathbf{x}}^{\Re}_{1}$ and $\check{\mathbf{x}}^{\Re}_{k}$. Note that $d \leq N_{\mathrm{t}}$ and $|\mathcal{I}_{\mathrm{c}}| = N_{\mathrm{t}} - d$  for BPSK signaling. 
The two vectors $\mathbf{g}_1^{\Re}$ and $\mathbf{g}_k^{\Re}$ can now be expressed as:
\begin{eqnarray}
\mathbf{g}_k^{\Re} &=& \mathbf{g}_{\mathrm{c}} + \mathbf{g}_{\mathrm{d}} \nonumber \\
\mathbf{g}_{k'}^{\Re} &=& \mathbf{g}_{\mathrm{c}} - \mathbf{g}_{\mathrm{d}}
\end{eqnarray}
where $\mathbf{g}_{\mathrm{c}}$ and $\mathbf{g}_{\mathrm{d}}$ are the summations of the $N_{\mathrm{t}}-d$ and $d$ columns of $\mathbf{H}$ corresponding to the indices given in $\mathcal{I}_{\mathrm{c}}$ and $\mathcal{I}_{\mathrm{d}}$, respectively. For Rayleigh fading with unit variance, $\mathbf{g}_{\mathrm{c}}$ is $ \mathcal{N}(\mathbf{0},\frac{N_{\mathrm{t}}-d}{2} \mathbf{I}_{2N_{\mathrm{r}}})$ and $\mathbf{g}_{\mathrm{d}} $ is $\mathcal{N}(\mathbf{0},\frac{d}{2} \mathbf{I}_{2N_{\mathrm{r}}})$.

The probability that $\operatorname{sign}(g^{\Re}_{1,i}) = \operatorname{sign}(g^{\Re}_{k,i})$ is given as
\begin{equation}
\mathbb{P}\big[\operatorname{sign}(g^{\Re}_{k,i}) = \operatorname{sign}(g^{\Re}_{k',i})\big] = \frac{2}{\pi}\operatorname{arctan}\sqrt{\frac{N_{\mathrm{t}}-d}{d}}.
\label{eq_sign_sign}
\end{equation}
This is obtained by applying a result in \cite{choi2016near}, which states that if $a \sim \mathcal{N}(0,\sigma_a^2)$ and $b \sim \mathcal{N}(0,\sigma_b^2)$ then 
\begin{equation}
\mathbb{P}\big[\operatorname{sign}(a + b) = \operatorname{sign}(a - b)\big] = \frac{2}{\pi}\operatorname{arctan}\frac{\sigma_a}{\sigma_b}.
\end{equation}
Due to the independence between the events $\operatorname{sign}(g^{\Re}_{k,i}) = \operatorname{sign}(g^{\Re}_{k',i})$, for $i=1,\ldots,2N_{\mathrm{r}}$, the result in \eqref{eq_g_1_equals_g_2} thus follows.

\section{Proof of Proposition \ref{proposition-2}} \label{append-D}
Without loss of generality, we assume that $\check{\mathbf{x}}_1^{\Re} = [\mathbf{1}_{N_{\mathrm{t}}}^T, \mathbf{0}_{N_{\mathrm{t}}}^T]^T$ was transmitted. Denote $E_k$, $1<k\leq K$, as the event $\check{\mathbf{y}}_{1} = \check{\mathbf{y}}_{k}$. The detection error event $E$ is then defined as $E = \bigcup_{k>1} E_k$. We want to find the VER given event $E$ and subsequently prove that $P^{\mathrm{ver}}_{\rho \rightarrow \infty}  \leq \frac{1}{2}\sum^K_{k>1}\mathbb{P}(E_k)$. We note that $E_2,\ldots,E_K$ are not necessarily mutually exclusive nor independent.  However, we can combine $E_2,\ldots,E_K$ into larger events $G_1,\ldots,G_L$ that are mutually exclusive. Herein, the rule for forming $G_{\ell}$ is as follows:
\begin{enumerate}
	\item If $E_k$ is mutually exclusive with all other events, then $E_k \subset G_1$.
	\item If a pair of events $E_k$ and $E_m$ intersect, i.e., $E_k\cap E_m\neq \varnothing$, but $E_k \cup E_m$ is mutually exclusive with all other events, then $(E_k \cup E_m) \subset G_2$.
	\item $G_3,\ldots,G_L$ are then formed in a similar fashion.
\end{enumerate}
Certainly, if $E_k \subset G_{\ell}$, then $E_k \cap G_{\ell'} = \varnothing$, for $\ell'\neq \ell$. This combining strategy effectively partitions $E$ into mutually exclusive events $G_1,\ldots,G_L$. The VER is calculated as:
\begin{enumerate}
	\item If event $E_k \subset G_1$ has occurred, the receiver would erroneously pick the detected vector $\hat{\mathbf{x}}^{\Re}_{k}\neq \check{\mathbf{x}}^{\Re}_1$ with a probability of $1/2$, i.e., $\mathrm{VER}= 1/2$.
	\item For any two events $E_k, E_m \subset G_2$ and $E_k \cap E_m \neq \varnothing$, we consider the following three partitions of $E_k\cup E_m$:
	\begin{itemize}
		\item If $E_k\cap E_m^{\mathrm{c}}$ has occurred, $\mathrm{VER}= 1/2$.
		\item If $E_k^{\mathrm{c}}\cap E_m$ has occurred, $\mathrm{VER}= 1/2$.
		\item If $E_k\cap E_m$ has occurred, the receiver would erroneously pick the detected vector as either $\hat{\mathbf{x}}^{\Re}_{k}$ or $\hat{\mathbf{x}}^{\Re}_{m}$ with a probability of $2/3$, i.e., $\mathrm{VER}= 2/3$.
	\end{itemize}
	We then have
	\begin{eqnarray}
	&&\frac{1}{2} \mathbb{P}[E_k\cap E_m^{\mathrm{c}}] + \frac{1}{2} \mathbb{P}[E_k^{\mathrm{c}}\cap E_m] + \frac{2}{3} \mathbb{P}[E_k\cap E_m]  \nonumber \\
	&\leq& \frac{1}{2} \mathbb{P}[E_k\cap E_m^{\mathrm{c}}] + \frac{1}{2} \mathbb{P}[E_k^{\mathrm{c}}\cap E_m] +  \mathbb{P}[E_k\cap E_m] \nonumber \\
	&=& \frac{1}{2}\mathbb{P}[E_k] + \frac{1}{2}\mathbb{P}[E_m].
	\end{eqnarray}
	\item The same principle of partitioning can be applied for events in $G_3,\ldots,G_L$ to calculate the VER.
\end{enumerate}
Therefore, $P^{\mathrm{ver}}_{\rho \rightarrow \infty}$ is upper-bounded as
\begin{eqnarray}
P^{\mathrm{ver}}_{\rho \rightarrow \infty} &\leq& \sum_{E_k\subset G_1}  \frac{1}{2} \mathbb{P}[E_k] + \sum_{E_k\subset G_2} \frac{1}{2}\mathbb{P}[E_k] + \ldots \nonumber \\
&=& \frac{1}{2} \sum_{k>1}^K   \mathbb{P}[E_k].
\end{eqnarray}
The inequality presented in the proposition follows by combining the result in Theorem \ref{theorem-2} and noting that there are $\binom{N_{\mathrm{t}}}{d}$ labels with Hamming distance $d$ from $\check{\mathbf{x}}_1^{\Re}$.

\ifCLASSOPTIONcaptionsoff
  \newpage
\fi

\bibliographystyle{IEEEtran}

\end{document}